\documentclass[twocolumn,apl,amsmath,amssymb,showpacs]{revtex4-1}
\usepackage{epsf}      
\usepackage{graphicx}
\usepackage{color}
\usepackage{soul}
\usepackage{gensymb}
\usepackage{sidecap}
\usepackage{amsmath}
\usepackage[hidelinks,colorlinks=true,linkcolor=blue,citecolor=blue]{hyperref}
\DeclareUnicodeCharacter{2212}{\textendash}

\begin{document}
\title{Mixed Polyanionic NaFe$_{1.6}$V$_{0.4}$(PO$_{4}$)(SO$_{4}$)$_{2}$@CNT Cathode for Sodium-ion Batteries: Electrochemical Diffusion Kinetics and Distribution of Relaxation Time Analysis at Different Temperatures}     
           
\author{Jayashree Pati}
\affiliation{Department of Physics, Indian Institute of Technology Delhi, Hauz Khas, New Delhi-110016, India}
\author{Rajendra S. Dhaka}
\email{rsdhaka@physics.iitd.ac.in}
\affiliation{Department of Physics, Indian Institute of Technology Delhi, Hauz Khas, New Delhi-110016, India}

\date{\today}      

\begin{abstract}
We report the electrochemical sodium-ion kiinetics and distribution of relaxation time (DRT) analysis of a newly designed mixed polyanionic NaFe$_{1.6}$V$_{0.4}$(PO$_{4}$)(SO$_{4}$)$_{2}$@CNT composite as a cathode. The specific capacity of 104 mAhg$^{-1}$ is observed at 0.1~C with the average working voltage of $\sim$3~V. Intriguingly, a remarkable rate capability and reversibility are demonstrated up to very high current rate of 25~C. The long cycling test up to 10~C shows high capacity retention even after 2000 cycles. The detailed analysis of galvanostatic intermittent titration technique (GITT) and cyclic voltammetry (CV) data reveal the diffusion coefficient of 10$^{-8}$--10$^{-11}$ cm$^{2}$s$^{-1}$. We find excellent stability in the thermal testing between 25--55$^\circ$C temperatures and 80\% capacity retention up to 100 cycles at 5~C. Further, we analyse the individual electrochemical processes in the time domain using the novel DRT technique at different temperatures. The {\it ex-situ} investigation shows the stable and reversible structure, morphology and electronic states of the long cycled cathode material. More importantly, we demonstrate relatively high specific energy of $\approx$155 Wh kg$^{-1}$ (considering the total active material loading of both the electrodes) at 0.2~C for full cell battery having excellent rate capability up to 10~C and long cyclic stability at 1~C. \\ 
\end{abstract}  .
\maketitle

\section{\noindent ~Introduction}

In recent years, the gradual expansion and development of renewable resources (i.e., wind, solar, geothermal and water tides) create a demand for cost-effective energy storage systems (EES)  for sustainable development \cite{YabuuchiCR14}. Now-a-days lithium ion batteries (LIBs) are widely used in making commercially applicable electric vehicles (EVs), portable devices, grid EES and proves as the most efficient battery technology in terms of high energy density and large power output \cite{ArmandNature08}. However, the issues of lithium extraction costs, non-uniform geographical distribution, scarcity in earth's crust (20 ppm)\cite{ArmandNature08} and hazardous effect of Li-metal with its electrolytes create many challenges for their wide applications in the future market. These challenges in LIBs shout out a need for alternative battery technology, which proves as a promising candidate for multi-billion battery industry. In this line, the sodium-ion batteries (SIBs) have been considered as a green and cost-effective solution due to its natural abundance, eco-efficient nature and similar electro-chemistry as that of LIB. Despite of these advantages, the SIBs suffer slow Na-ion kinetics, low specific energy and power density and cycling issues owing to its high molecular weight (23 g mol$^{-1}$), higher redox potential (-2.71~V vs. SHE) and larger radius (1.02~\AA) as compared to the lithium \cite{VaalmaNRM18, LiNE22}. To overcome with some of these challenges faced in SIBs, it is vital to design new composite electrode materials, understand the diffusion kinetics and target for high energy density and long cyclic life along with fast charging capability and the thermal stability. Here, the layered metal oxides and polyanionic compounds drag profound attention as potential cathodes \cite{GaoESM20, YouAEM18, LiAFM20, BarpandaNC14, SapraWEE21, LiueS24} where the layered oxides exhibit high specific capacity and good electronic conductivity, but possess low operating potential, poor cycle stability, irreversible phase transitions and highly hygroscopic nature. On the other hand, some recently developed high entropy layered oxides helps to improve the energy density and cycle life by curbing the irreversible phase transitions \cite{TianJMCA22, DangASS23}. 

Among various complex oxides explored in past few decades, the NASICON (Na Super Ionic CONductor) type polyanionic based structures are considered suitable cathode materials for high performance energy storage devices due to their high ionic conductivity, high operating voltage, very stable framework, low gravimetric and volumetric capacity, diverse cationic and anionic substitutions with tunable characteristics \cite{YouAEM18, LinCEC22}. In general, the polyanionic structure involves tetrahedral anions (XO$_{4}$)$^{n-}$ or their derivatives ({\it X}$_{m}$O$_{3m+1}$)$^{n-}$ ($X=$ P, S, Si, B, W or Mo) connected with {\it M}O$_{x}$ polyhedra ($M=$ transition metal) \cite{LiAFM20}. This characteristic structure provides stability during the redox reactions and enhance the ionic conductivity. The operating potential of these materials can be tuned through inductive effect where the $X-$O and $M-$O bonds in polyhedra play an important role \cite{YouAEM18, BarpandaNC14, SapraWEE21}. More interestingly, the mixed polyanionic compounds possess robust and durable structural support for continuous de-/intercalation of Na-ion with relatively high redox potential owing to their inductive effect of high electro-negative anionic groups \cite{SapraWEE21}. For example, a NASICON type mixed polyanionic Na$_{4}$Fe$_{3}$(PO$_{4}$)$_{2}$P$_{2}$O$_{7}$ (NFPP) cathode material delivers a theoretical specific capacity of 129 mAhg$^{-1}$ and working voltage of 3.2~V with multiple redox activity centres \cite{KimJACS12}. In fact, the multi electron reaction in carbonophosphate Na$_{3}$MnPO$_{4}$(CO$_{3}$) provides high specific energy 374 Wh/kg \cite{ChenCM13}. Also, the Na$_{4}$Co$_{3}$(PO$_{4}$)$_{2}$(P$_{2}$O$_{7}$) delivers a bit less specific capacity of 95 mAhg$^{-1}$ at 0.2~C, but a very high working voltage upto 4.5~V due to its inductive effect and stable crystal structure \cite{NoseJPS13}. This material shows anodic characteristic \cite{DwivediACSAEM21} and paves the way towards its use in symmetric cells. 

A useful approach is to combine the high electronegative nature of (SO$_{4}$)$^{2-}$ with the PO$_{4}$ group, which for the first time was reported by Goodenough's group on a high-voltage phospho-sulfate cathode NaFe$_{2}$(PO$_{4}$)(SO$_{4}$)$_{2}$ \cite{ShivaEES16}. Then, Belharouak's group provided a detailed electrochemical investigation on this cathode material, which shows a specific capacity of 89 mAhg$^{-1}$ at 0.05~C having a working voltage of 3~V \cite{YahiaJPS18}. Li {\it et al.} modified this material with rGO, enhancing the specific capacity upto 90 mAhg$^{-1}$ at 25 mAg$^{-1}$ (0.2~C) and also maintains upto $\approx$60\% retention after 300 cycles tested at 50 mAg$^{-1}$ \cite{LiNJC21}. Further, the V doping at the Fe site, i.e., the optimized NaFe$_{1.6}$V$_{0.4}$(PO$_{4}$)(SO$_{4}$)$_{2}$ cathode exhibits the specific capacity of around 90 mAhg$^{-1}$ at 0.1~C and a retention of 96\% is reported after 50 cycling, which was found to be better than other compositions \cite{EssehliJPS20}. Additionally, the calculated diffusion coefficients of this material from GITT falls in the range of 6--7$\times$10$^{-11}$ cm$^{2}$s$^{-1}$ \cite{EssehliJPS20}. However, for in depth understanding it is important to perform the advanced analysis including DRT (Distribution of Relaxation time) and temperature dependent behavior for these types of polyanionic materials \cite{ChenNC19, LanBS21, LiCAEJ23, LiangAMI17}. Intriguingly, the DRT analysis at different charged states depict the polarization resistances at different time constants \cite{LanBS21, LiCAEJ23}. Also, the DRT analysis can provide direct information on the number of individual processes occurring due to the electro-chemical reactions and their relative contribution to the total impedance without fitting the large number of equivalent circuits \cite{IurilliIEEE22}. Further, the thermal behavior of Na$_3$V$_2$(PO$_4$)$_3$ at 55$^\circ$C showed the capacity retentions of 50.1\% and 24.7\% after 1000 cycles at 5~C. The electrochemical study on rGO modified NFPP at 50$^\circ$C was reported to exhibit a significant capacity retention of 91.4\% at 0.5~C \cite{ChenNC19}. 

Therefore, we investigate a detailed electrochemical analysis including DRT and thermal stability of NASICON type NaFe$_{1.6}$V$_{0.4}$(PO$_{4}$)(SO$_{4}$)$_{2}$@CNT (NFVPS@CNT) composite, which shows the high rate capability, long cycling and thermal stability. The optimized concentration of vanadium doping and the inter-connected channels of CNT increase the electronic conductivity and vacancy defect sites, which demonstrated an  improved specific discharge capacity of 104~mAhg$^{-1}$ at 0.1~C with operating voltage of 3~V. The severe cycling of cells upto very high current rate of 25~C with 95\% retention of initial capacity upon cycling back to 0.1~C signifies the highly reversible nature of this cathode material. Intriguingly, the long cycling test at various current rates up to 10~C shows 71, 57 and 36\% capacity retention after 500, 1000, and 2000 cycles, respectively. The bulk diffusivity of Na-ion is evaluated through detailed analysis of GITT and CV, which falls in 10$^{-8}$ --10$^{-11}$ cm$^{2}$s$^{-1}$ range. Through these electrochemical testing, we demonstrate higher specific capacity, stability, reversibility and diffusivity of NFVPS@CNT composite as compared to the reports in refs.~\cite{ShivaEES16, YahiaJPS18, LiNJC21, EssehliJPS20}. More importantly, we used the novel DRT technique for this cathode material at different temperatures to analyse the individual electrochemical processes in the time domain. The thermal treatment of half-cells between 25--55$^\circ$C temperatures and long cycling at 45$^\circ$C holds out-standing capacity retention of 80\% upto 100 cycles at 5~C rate. Furthermore, the {\it ex-situ} XRD, FE-SEM and XPS analysis compares the phase, morphology and electronic states respectively, of the cycled and pristine material, which confirms the stable Na-ion dynamics and reversibility of NFVPS@CNT cathode.  \\

\section{\noindent ~Experimental Details:}

\subsection{\noindent ~Synthesis of NFVPS@CNT composite material:}

The NFVPS@CNT composite material was synthesized through sol-gel route where we use precursors ($\geq$99\% purity) in stoichiometric ratios of NaNO$_{3}$, NH$_{4}$VO$_{3}$, Fe(NO$_{3}$)$_{3}$.9H$_{2}$O, citric acid (C$_{6}$H$_{8}$O$_{7}$), NH$_{4}$H$_{2}$(PO$_{4}$)  and (NH$_{4}$)$_{2}$SO$_{4}$  for the reaction. The citric acid and NH$_{4}$VO$_{3}$ were added in 20 ml of distilled water in a mole ratio of 2:0.4 forming an orange colour solution. In order to modify the composite, 5 wt\% of the assumed product weight of CNT was added in 15 ml of water and ultra-sonicated for 30-40 mins for uniform dispersion, attachment of functional groups at the side walls and opens the convoluted ends of CNT. The NaNO$_{3}$, Fe(NO$_{3}$)$_{3}$.9H$_{2}$O and (NH$_{4}$)$_{2}$SO$_{4}$ were dissolved in the above orange solution with continuous stirring  by forming dark brown solution (solution-1). The NH$_{4}$H$_{2}$(PO$_{4}$) was added in 10 ml of distilled water to form solution-2. Then, solution-1 and CNT solution were gradually added in solution-2 with continuous stirring. The final solution was sonicated again for 1 hr at room temperature and then stirred at 80\degree C for overnight. The obtained powder was annealed under Argon flow at 400\degree C for 26 hrs to remove the volatile H$_{2}$O, NH$_{3}$ and NO$_{2}$ molecules, which finally formed as uniform NASICON phase. 

\subsection{\noindent ~Structural and physical characterizations:}

 In order to study the crystallographic structure of the synthesized material, the room temperature x-ray diffraction (XRD) patteren (using Panalytical Xpert 3 system) is recored with CuK$\alpha$ radiation (1.5406 \AA) in 2$\theta$ range of 10 to 80$^\circ$ at scan rate 2$^\circ$/min. The high-resolution transmission electron microscopy (HR-TEM) measurements are performed with a Tecnai G2 20 system to analyse the morphology and crystallinity. To investigate the stoichiometry and elemental distribution across the composite, we use EDX from RONTEC system model QuanTax 200. The Raman spectrum is recorded using a Renishaw inVia confocal Raman microscope 2400 lines/mm grating at a wavelength of 532 nm with a laser power of 10 mW on the sample. To probe the electronic states of individual elements in pristine sample as well as in charge/discharge states, we perform x-ray photo-electron spectroscopy (XPS) measurements using AXIS Supra system having monochromatic Al K$\alpha$ source (1486.6~eV) and energy resolution of ~0.5~eV. The  binding energy of C 1$s$ core-level at 284.6~eV is used to calibrate individual peak positions. We use the Voigt function and Tougaard method for deconvolution of the core-level peaks and background subtraction, respectively. 

\subsection{\noindent ~Coin cell fabrication:} 

The as-synthesized NFVPS@CNT composite is used as the cathode to fabricate the half-cells of SIBs to investigate the electrochemical performance. The active material, acetylene black (to improve electronic conductivity) and polyvinyldiuoride (PVDF) binder are mixed in N-methylpyrrolidone (NMP) solvent in weight ratio of 7:2:1 to form uniform slurry. The as prepared slurry was coated on a aluminium foil using a doctor blade method with active material mass loading about 2 mg cm$^{-2}$ followed by vacuum drying at 120$^{0}$C over night to evaporate the excessive solvent and moisture. The electrodes were cut and dried under vacuum before inserting in glove box (MBRAUN). The CR2032 type coin cells were fabricated in Ar filled glove box under controlled level of O$_2$ and H$_2$O ($<$0.1~ppm). The Na foil was used as the counter and reference electrode. We use 1 M NaPF$_6$ in a mixture of ethylene carbonate (EC) and diethyl carbonate (DEC) 1:1 (vol \%) as electrolyte. The experimental details of hard carbon testing as anode and the NFVPS@CNT\textbar\textbar HC full cell are provided in \cite{SIfile}. 

\subsection{\noindent ~Electrochemical and DRT analysis:}

The galvanostatic charge-discharge (GCD) profiles were measured by using a Neware battery analyzer BTS400 in a voltage window of 1.5--4.5 V (vs Na$^{+}$/Na) at different current rates. The cyclic voltammetry (CV) was conducted using a Biologic VMP-3 model in the same voltage range at different scan rates. The electrochemical impedance spectroscopy (EIS) measurements were performed using VMP-3 in the frequency range of 0.1~Hz to 100~kHz, and ac amplitude of 10 mV was used at the open circuit voltage of the cells. The DRT (Distribution of Relaxation Time) analysis of the corresponding EIS spectra was done using an open source MATLAB based program (DRT Tools). In order to fit the discrete impedance data, Tikhonov regularization was implemented with non-linear least-square method by fixing the parameter at 0.0001 with second-order regularization derivative. Further, the FWHM of the radial basis function (RBF) was set at 0.5 for getting appropriate fitting of the EIS data. After obtaining the DRT results, the peaks were fitted with Gaussian non-linear curve fitting method by using the Levenberg-Marquardt iteration in ORIGIN software to evaluate the values of time constant and polarization resistance. \\

\begin{figure*}
\includegraphics[width=6.5in]{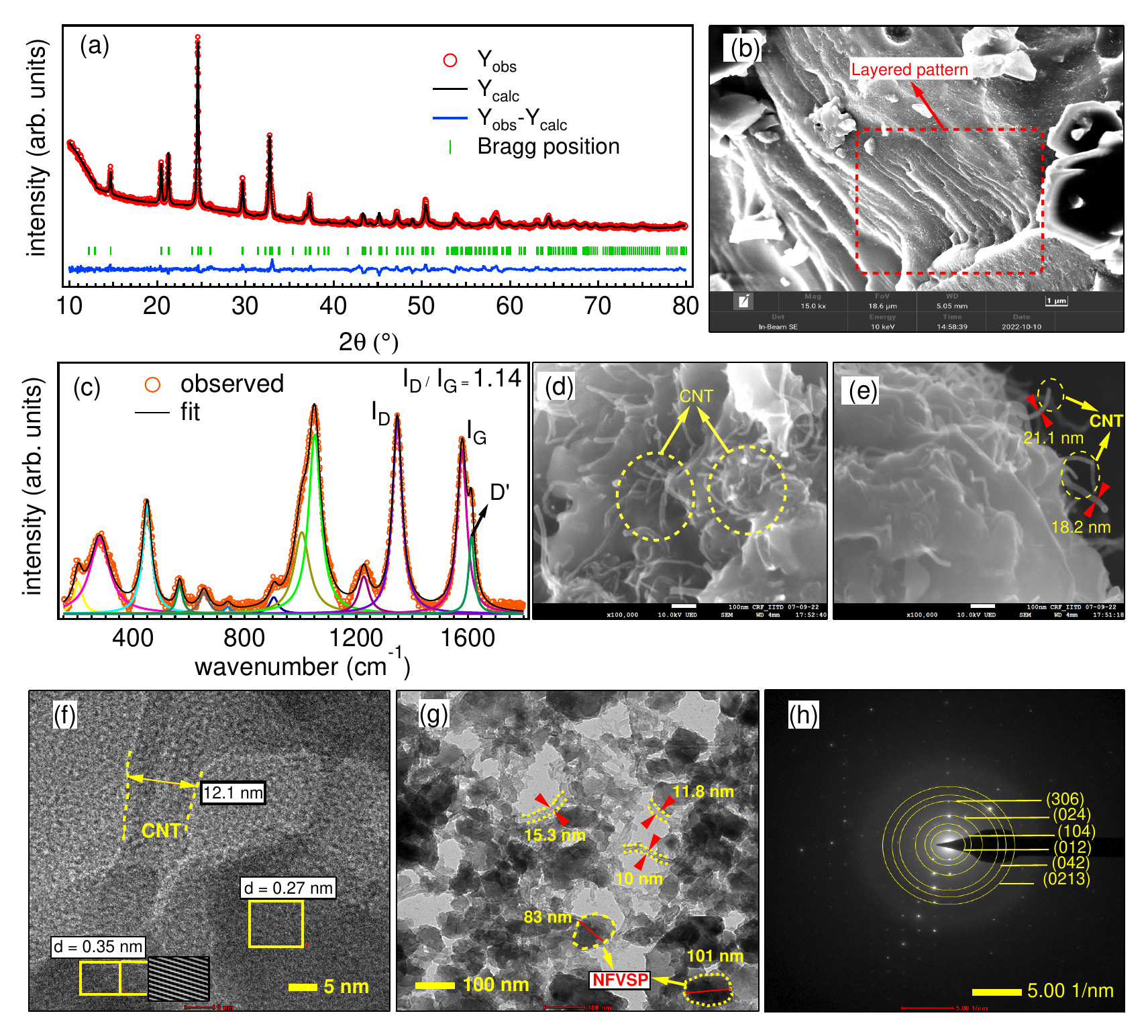}
\caption {{\bf Basic charectrizations of as prepared NaFe$_{1.6}$V$_{0.4}$(PO$_{4}$)(SO$_{4}$)$_{2}$@CNT sample:} (a) The Rietveld refinement (black color) of x-ray diffraction pattern (red color) with Brag peaks and difference curve, (b) the FE-SEM image showing the layered structure, (c) the Raman spectrum in a range of 100-2000 cm$^{-1}$, (d-e) the FE-SEM images showing embedded CNTs. The HR-TEM images showing 2-D lattice fringes with inter-planar distance and CNTs in (f). The well interconnected CNTs anchored on NFVPS particles and the SAED pattern with indexed planes are shown in (g) and (h), respectively.}
 \label{Fig.1}
\end{figure*}

\section{\noindent ~Results and discussion:}

\subsection{\noindent ~Structural and morphological characterization:}

The x-ray diffraction (XRD) pattern of the NFVPS@CNT material with Rietveld refinement and Bragg's positions are shown in Fig.~1(a). The refined pattern represents trigonal crystal symmetry (space group: {\it R-3c}, no.\#167) \cite{EssehliJPS20} with the lattice parameters  $a=b=$ 8.3536\AA, $c=$ 21.702\AA, $\alpha$ = 90\degree, $\beta$ = 90\degree and $\gamma$ = 120\degree and the values of R$_{p}$ (3.36\%) and R$_{wp}$ (4.49\%) define a good accordance of calculated pattern with the observed profile. The FE-SEM image in Fig.~1(b) depicts the layered morphology, which enhances the charge storage ability by providing large active surface area for the electrochemical reaction. In order to know the existence of CNT, phosphate and sulfate groups, the de-convoluted Raman spectrum is presented in Fig.~1(c). The most intense peak around 1004 cm$^{-1}$ corresponds to the symmetric stretching ($\nu$$_{1}$) vibrational mode of (SO$_{4}$)$^{2-}$ tetrahedra \cite{QiuIEE19, AngelAS12} and the peaks near 1000 cm$^{-1}$ attribute to the stretching vibrations of $\nu$$_{P-O-P}$ of (PO$_{4}$)$^{3-}$ anion \cite{MarkevichJPS11}. The peak at 1050 cm$^{-1}$ denotes the anti-symmetric internal stretching ($\nu$$_{3}$) vibration of PO$_{4}$ group \cite{WuNano13} and the bands observed at 737 cm$^{-1}$ attribute to the $\nu$$_{2}$ bending mode of (PO$_{4}$)$^{3-}$ anion, whereas the peaks between 400 to 600 cm$^{-1}$ comprise $\nu$$_{4}$ bending modes of PO$_{4}$ and SO$_{4}$ groups \cite{QiuIEE19, MarkevichJPS11}. All the other small structural vibrations below 400 cm$^{-1}$ are assigned to the Fe-O and V-O based lattice modes, which are reported to be sensitive to the metal substitution \cite{DifiJPCC15, CriadoFP19}. The Raman mode around 1200 cm$^{-1}$ depicts the characteristic features of MWCNTs (multi-walled CNTs), which include the D-band indicating A$_{g}$ mode at 1348 cm$^{-1}$ due to the introduction of disorder in the CNT structures. Also, the G-band at 1579 cm$^{-1}$ refers the E$_{2g}$ vibration mode owing to the in-plane displacement of carbon atoms strongly attached with the hexagonal structures \cite{GaoJMC17}. Additionally, a D-band appears at 1612 cm$^{-1}$ denoting the structural defects due to the amorphous carbonaceous materials \cite{GaoJMC17, ChoiJN13, MelvinJMC14}. Herein, the I$_{D}$/I$_{G}$ ratio of the composite is found to be 1.14, which indicates successful introduction of structural dis-orderness in the composite. This structural disorder is expected to increase the electronic conductivity of the composite, which reflects by having high-rate capability of the cathode material during the electrochemical reaction.

\begin{figure*}
\includegraphics[width=6.4in]{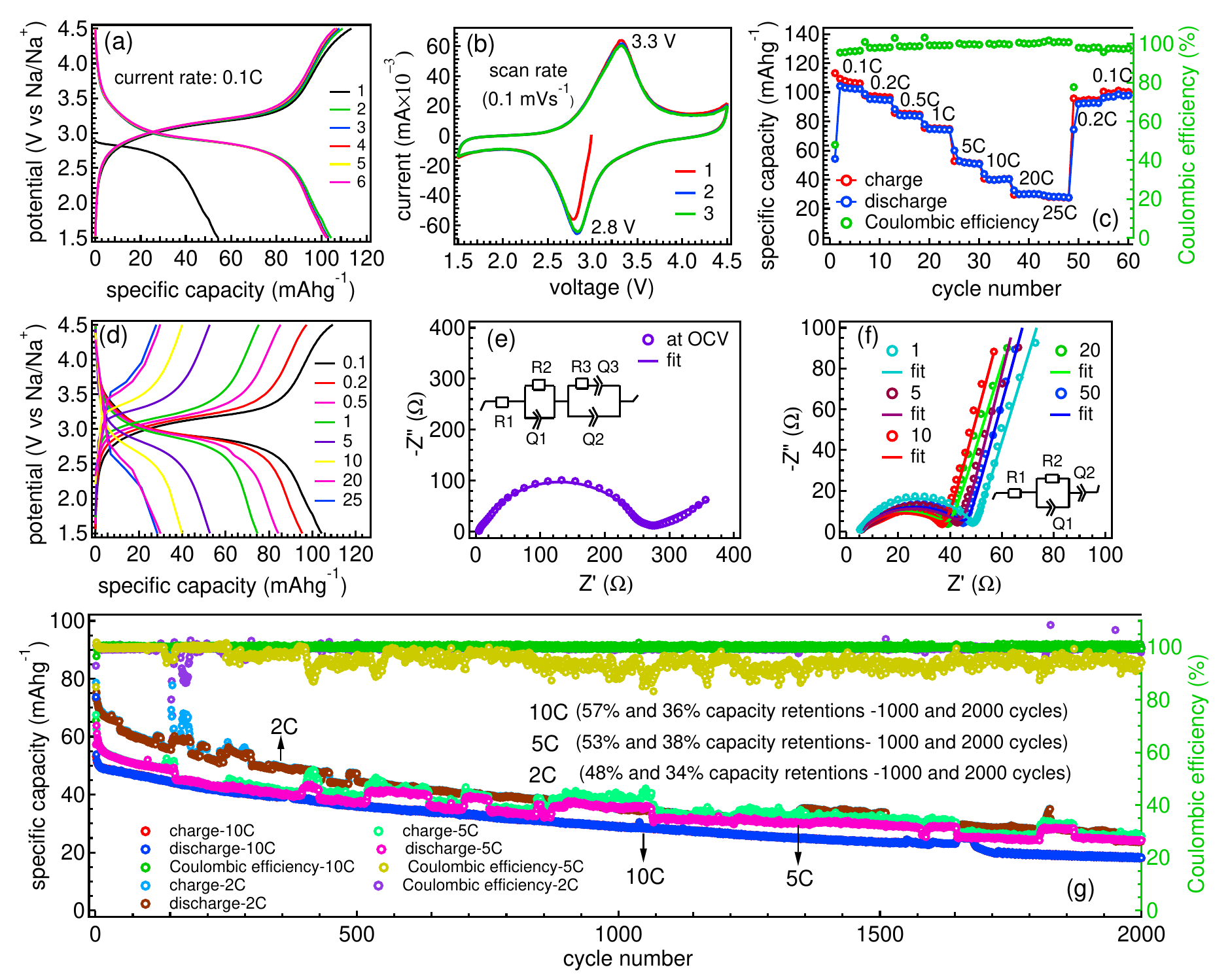}
\caption {{\bf Electrochemical measurements of NaFe$_{1.6}$V$_{0.4}$(PO$_{4}$)(SO$_{4}$)$_{2}$@CNT cathode:} (a) the galvanostatic discharge-charge voltage profiles at a current rate of 0.1 for 6 cycles, (b) the cyclic voltammetry curves at a scan rate of 0.1 mVs$^{-1}$ for 3 cycles in a voltage window of 1.5--4.5~V, (c) the rate capability measurement at different current rates for 6 cycles each in a voltage window of 1.5--4.5~V, (d) the galvanostatic discharge-charge voltage profiles of 2$^{nd}$ cycles at different current rates from 0.1 to 25~C, (e) the EIS spectrum of freshly prepared half-cell (at OCV), (f) the EIS spectra of the same half cell as presented in (f) after 1$^{st}$, 5$^{th}$, 10$^{th}$, 20$^{th}$ and 50$^{th}$ cycles along with circuit diagram, (g) the specific capacity and Coulombic efficiency (\%) in long cycling measurement at different current rates of 2~C, 5~C and 10~C up to 2000 cycles.}
 \label{Fig.2}
\end{figure*}

Figs.~1(d, e) show very well structural integration between NFVPS and CNT, where the successful incorporation of CNT clusters within the NFVPS bulk structure owing to the tight binding of CNTs (indicated by dashed yellow circles) with the NFVPS crystal structure. These interconnected and firmly anchored CNTs on NFVPS particles provide high electronic conductivity resulting superior performance in the rate capability, discussed later. In Figs.~1(f--h), the high-resolution transmission electron microscopy (HR-TEM) confirms the well crystalline nature of the material, as clearly demonstrated from the 2-D lattice fringes mentioned in Fig.~1(f). The inter-planar distances of 0.35 nm and 0.27 nm are correspond to the planes (006) and (116), respectively. The diameter of the CNTs varies between 10 and 21~nm [see Figs.~1(e--g)], which are homogeneously dispersed between the aggregated NFVPS material, see Fig.~1(g). In addition, the SAED (selected area electron diffraction) pattern shown in Fig.~1(h) depicts the indexed bright crystal planes (012), (104), (024), (306), (042) and (0213) corresponding to the inter-planar distance of 0.62, 0.44, 0.30, 0.20, 0.17 and 0.15~nm, respectively. 

\subsection{\noindent ~Electrochemical analysis:} 

In order to get a comprehensive idea about the charge storage activity and redox mechanisim of the NFVPS@CNT composite material, the galvanostatic charging discharging (GCD) measurements are performed at a current rate of 0.1~C in a voltage window of 1.5--4.5~V, as presented in Fig.~2(a) up to six cycles. The half-cell was first discharged to 1.5~V to convert Fe$^{3+}$ to Fe$^{2+}$ by reducing the material and again charging it back to higher cut-off voltage up to 4.5~V. We observe a significantly high specific discharge capacity of 104 mAhg$^{-1}$ in the 2$^{nd}$ initial cycle at 0.1~C with a capacity retention of 98\% up to 6$^{th}$ cycle. The observed plateau and maximum specific capacity are consistent with 1.6 electron reaction, which is considered to be 1~C = 102 mAg$^{-1}$. This high capacity is possible due to the vacancy defects created from the vanadium doping and CNT-modification, which improved the electronic conductivity \cite{ChandelDT22}. In this context, the V$^{5+}$ doping at the Fe$^{3+}$ site can produce oxygen vacancies with a partial reduction of Fe$^{3+}$ to Fe$^{2+}$, which also helps to enhance the electronic and ionic conductivity due to the higher ionic radius of Fe$^{2+}$ as compared to the Fe$^{3+}$ \cite{ChandelDT22}. As a result, the activation barrier for Na-ion migration decreases, providing ease insertion/de-insertion of Na-ion in the bulk structure. The CV measurement in Fig.~2(b) depicts the activation process of the Fe$^{3+/2+}$ redox couple during Na-insertion/de-insertion in a voltage window of 1.5--4.5~V at a scan rate of 0.1 mVs$^{-1}$ for 3 cycles \cite{ShivaEES16, YahiaJPS18}. Here, the initial insertion process of the 1$^{st}$ cycle shows the conversion of less number of Fe$^{3+}$ to Fe$^{2+}$ resulting lower redox current, which is consistent with the GCD profile, attributing low initial specific discharge capacity (54 mAhg$^{-1}$) in the 1$^{st}$ discharge [Fig.~2(a)]. After that, the capacity is found to be stable around 104 mAhg$^{-1}$. 

Fig.~2(c) presents the rate capability measurements at different current rates up to 25~C with the Coulombic efficiency nearly 100\% except the 1$^{st}$ cycle. The NFVPS@CNT electrode delivers maximum specific discharge capacities of 104, 99, 88.3, 78, 60, 44, 33 and 29 mAhg$^{-1}$ at current rates of 0.1, 0.2, 0.5, 1, 5, 10, 20 and 25~C respectively, for 6 cycles each. After testing the cell up to 25~C, we again checked the capacity at 0.1~C and found 95\% retention of the initial capacity. These rate capability measurement assures the highly reversible nature and more importantly fast charging capability of this electrode, which are crucial to improve the power density factor of the battery. The corresponding GCD profiles of 2$^{nd}$ cycles of Fig.2~(c) at different current rates are provided in Fig.~2(d), indicating single-phase behavior during charge/dis-charge with the operating voltage of $\sim$3~V. The material is also tested in full cell configuration using hard carbon as anode, see Fig.~S5 of ref.~\cite{SIfile}. Moreover, we use the electro-chemical impedance spectroscopy (EIS), i.e., the Nyquist plot at OCV to understand the Na-ion transport mechanism during cycling, where the fitted circuit element R1 attributes the solution resistance of the electrolyte and intrinsic resistance, the two parallely connected constant phase element (Q) and resistance (R) determine the contribution of SEI layer and charge transfer between SEI/electrode interface, as shown in Fig.~2(e). In Fig.~2(f), we present the EIS plots measured after 1$^{st}$, 5$^{th}$, 10$^{th}$, 20$^{th}$ and 50$^{th}$ cycles. We note that the charge transfer resistance gradually decreases from 1$^{st}$ to 10$^{th}$ cycle due to the degradation of SEI layer and activation process. After 10$^{th}$ cycle, the impedance increases for 20$^{th}$ and 50$^{th}$ cycle indicating large over-potential due to the electrolyte degradation in products at high voltage.The long cycling measurement in Fig.~2(g) records 48\%, 53\% and 57\% capacity retentions between 3$^{rd}$ and 1000 cycles at 2~C, 5~C and 10~C respectively; whereas after 2000 cycling the retention is found to be 34\%, 38\%, and 36\% at 2~C, 5~C and 10~C respectively. Also, we observe some fluctuations in specific capacity and decrease in Coulombic efficiency at 2~C rate, which might be due to the lack of stability of SEI layer at higher potentials above 4.2~V \cite{GuanMAT21} as well as the effect of relaxation time/state and internal temperature of cell during long cycling \cite{ChoiPCCP18}. After these measurement, the cell cycled at 10~C was dismantled and the cathode material was procured for XRD and FE-SEM measurements to study the structural and morphological changes during the cycling, as shown in the Figs.~S2(a, b) of ref.~\cite{SIfile}.  

\begin{figure*} 
\includegraphics[width=6.9in]{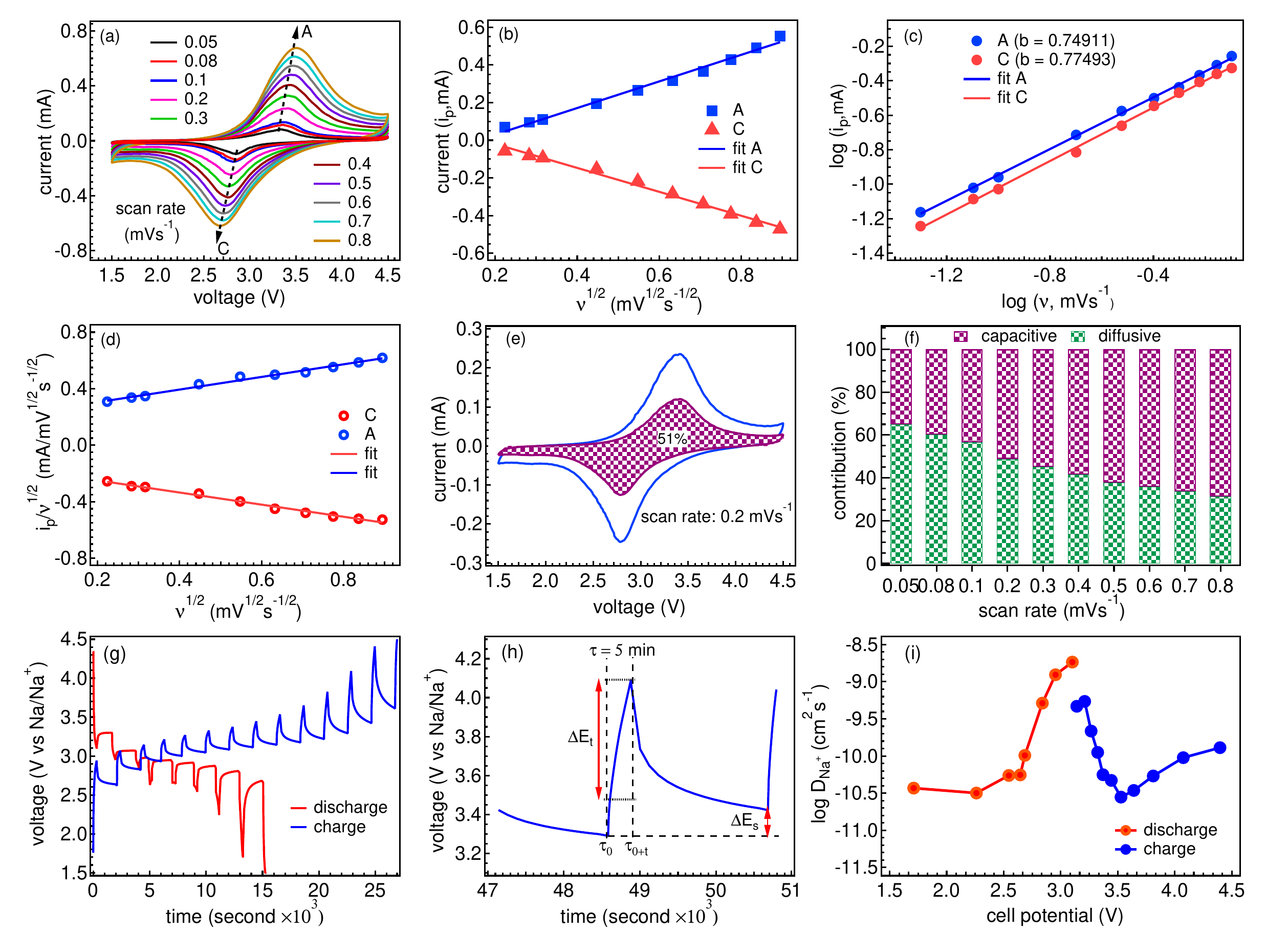}
\caption {{\bf Quantitative and kinetic analysis of Na$^{+}$ diffusion coefficient in NaFe$_{1.6}$V$_{0.4}$(PO$_{4}$)(SO$_{4}$)$_{2}$@CNT cathode:} (a) The cyclic voltammetry curves at different scan rates in a voltage window of 1.5--4.5~V, having anodic (A) and cathodic (C) peaks, (b) the linear relation between peak current $i_{p}$ and square root of scan rate for different scan rates, (c) the linear fitting between log $i$ versus log $\nu$, (d) the linear relation plot between $i(V)$/$\nu$$^{1/2}$ versus $\nu$$^{1/2}$ for both cathodic and anodic peaks, (e) the capacitive contributed part for 0.2 mVs$^{-1}$ shown by shaded area, (f) the contributions of capacitive and diffusive current at different scan rates, (g) the GITT measurements at a current rate of 0.5~C in a voltage window of 1.5--4.5~V for a fresh cell, (h) the representation of different parameters in a single titration curve before, during and after application of a current pulse for 5 min, (i) the logarithmic variation of diffusion coefficient as a function of cell voltage for the GITT profile shown in (g).}
 \label{CV}
\end{figure*}

\subsection{\noindent ~Cyclic voltammetry and GITT analysis:} 

In order to study the redox processes in the fabricated half cells, we use the cyclic voltammetry (CV) technique, which also probes the chemical reactions initiated by electron transfer. Fig.~3(a) shows the CV curves at different scan rates for 2$^{nd}$ cycle each in a voltage window of 1.5--4.5~V. The anodic and cathodic peaks are denoted as A and C, respectively and determine the reversible redox reaction of Fe $^{3+/2+}$ occurring in both forward and reverse scans. The peak height and area gradually increases with the scan rate, indicating the occurrence of more number of redox reactions at higher scan rates. This is consistent as the peak area divided by the scan rate results the capacity, which should be constant. Notably, the redox peaks inherited the original shape even at higher scan rate of 0.5 mVs$^{-1}$ with a small shift, indicating small polarization effect in the NFVPS@CNT cathode material. In Fig.~3(b), the linear relationship between peak current (i$_{p}$) and square root of scan rate ($\nu$$^\frac{1}{2}$) determines the slope for the calculation of diffusion coefficient. If we consider the Na-ion diffusion as the rate limiting step and the charge transfer resistance at the interface is very negligible, then the peak current follows a linear relationship with $\nu$$^\frac{1}{2}$, as per the equation below \cite{DwivediACSAEM21}: 
\begin{eqnarray}
 i_{p} = (2.69 \times 10^{5}) A D^{\frac{1}{2}}  C n^{\frac{3}{2}} \mathcal{V}^{\frac{1}{2}}
 \end{eqnarray}
Using the above Randles-Sevcik equation, we calculate the diffusion coefficient of Na-ion during the insertion/extraction, where the $i_{p}$ is termed as the peak current (mA), $\nu$ is the scan rate (mVs$^{-1}$), {\it A} is the active surface area of the electrode (cm$^{2}$), $n$ is the number of electron transferred during the chemical reaction, {\it C} is the initial concentration of Na-ion in the electrode (mol/cm$^{3}$) and {\it D} is the diffusion coefficient of Na-ion in the electrode material (cm$^{2}$s$^{-1}$). Considering the appropriate values of these parameters, the calculated values of {\it D} for anodic and cathodic peaks are found to be in the range of 2.6$\times$10$^{-8}$ to 3.2$\times$10$^{-10}$cm$^{2}$s$^{-1}$. 

Moreover, in order to quantitatively demonstrate about the surface controlled (pseudo-capacitive) and/or Faradic nature of the cathode material, we perform detailed analysis of CV data using the power law \cite{PatiJMCA22}:
\begin{eqnarray}
i_{p} = a\nu^{b}
\end{eqnarray}
here, the $i_{p}$ determines the peak current value at the corresponding scan rate, $a$ and $b$ are the parameters. We can obtain equation (3) by taking the logarithmic function of equation (2) as described below \cite{SinghCEJ23}:
\begin{eqnarray}
log~i = b~log~\nu + log~a
\end{eqnarray}
where, {\it b} is the slope obtained from the linear fitting of log $i_{p}$ vs log $\nu$ plot, as shown in Fig.~3(c), which attributes the pseudo-capacitive behavior. The Faradic nature of the material dominates, when the value of {\it b} = 0.5 and {\it b} = 1 determines the capacitive behavior. The value of {\it b} in between 0.5 and 1 suggest for the pseudo-capacitance contribution (both surface controlled and diffusion controlled) in the electrode material. In the present case, the obtained values of {\it b} for anodic and cathodic peaks are 0.74 and 0.77 respectively, which represent the pseudo-capacitive nature of the NFVPS@CNT cathode material. 
 
 \begin{figure*}
\includegraphics[width=6.8in]{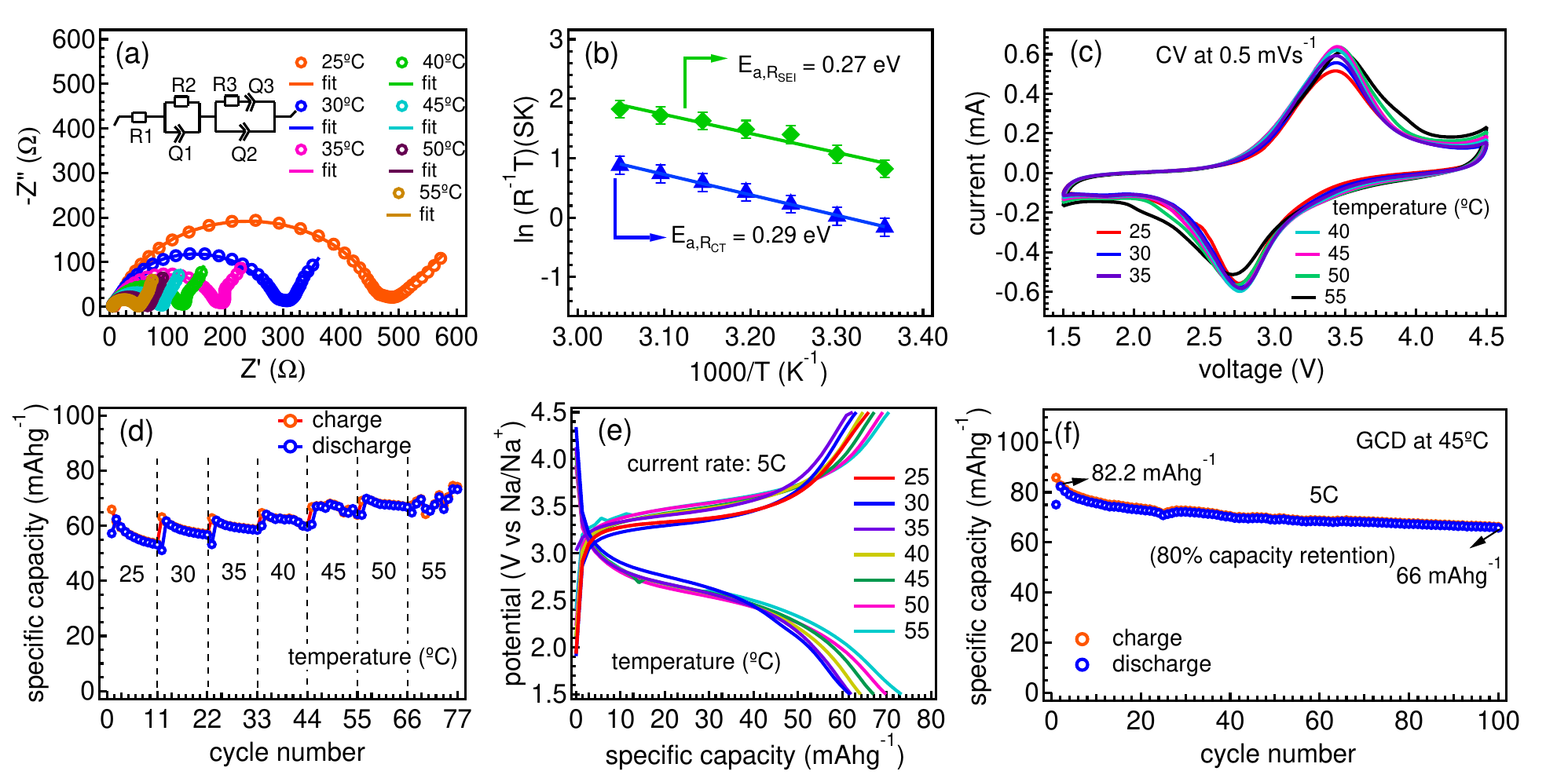}
\caption {{\bf The electrochemical performance of NaFe$_{1.6}$V$_{0.4}$(PO$_{4}$)(SO$_{4}$)$_{2}$@CNT cathode at different temperatures:} (a) The EIS spectra recorded at  25$^\circ$C, 30$^\circ$C, 35$^\circ$C, 40$^\circ$C, 45$^\circ$C, 50$^\circ$C and 55$^\circ$C, i.e., the fitted Nyquist plots with equivalent circuit in the inset, (b) the Arrhenius plot of the resistance contributions of the solid electrolyte interface R$_{SEI}$ and the charge transfer resistance R$_{CT}$, (c) the cyclic voltammetry at a scan rate of 0.5 mVs$^{-1}$. The galvanostatic charge-discharge specific capacity (d) measured at a current rate of 5~C up to 11 cycles at each temperature indicated, (e) the corresponding galvanostatic charge-discharge profiles of 2$^{nd}$ cycles of (d) at different temperatures from 25--55 $^\circ$C at 5~C current rate, and (f) the long cycling up to 100 cycles measured at 5~C current rate and 45$^\circ$C temperature.}
 \label{EIS}
\end{figure*}
 
Now it is important to get more insight about the pseudo-capacitance nature, i.e., to quantify the capacitive controlled and diffusion controlled reactions at a particular voltage using the equation below \cite{PatiJMCA22, SinghCEJ23}: 
\begin{eqnarray}
i(V) = k_{1} \nu+k_{2} \nu^{1/2}
 \end{eqnarray}
here, the $i(V)$ corresponds to the total current response from both capacitive and diffusive part, {\it k}$_{1}$$\nu$ depicts the capacitive effects and {\it k}$_{2}$$\nu$$^{1/2}$ represents the diffusion controlled effects. The constants k$_{1}$ and k$_{2}$ are resulted from the linear relation between {\it i(V)}/$\nu$$^{1/2}$ and $\nu$$^{1/2}$, as  below: 
\begin{eqnarray}
[i(V)/\nu^{1/2}] = k_{1} \nu^{1/2}+k_{2}
 \end{eqnarray} 
where {\it k}$_{1}$ represents the slope and {\it k}$_{2}$ determines the intercept of the linear relation, as shown in Fig.3~(d). More clearly, the shaded region in Fig.3~(e) represents the capacitive controlled part, which found to be around 51\% of the total current contribution. The area of simulated curve is found to be slightly more than the actual area of the CV curves, which can be due to small residual current in the charge storage mechanism \cite{PuACI21}. Moreover, we observe a clear increase of the capacitive contribution with the scan rate, see Fig.~3(f). Note that the diffusive contribution dominates at slow scan rate like 0.05 mVs$^{-1}$ providing 65.2\% of the total contribution, which gradually decreases to 31.3\% at high scan rate of 0.8 mVs$^{-1}$. This trend indicates the dominance of surface controlled reactions at higher scan rates. In order to calculate the self-diffusivity of Na-ion in the host cathode material, the GITT measurements are performed for the first cycle at particular voltages at a fixed current rate of 5~C. Fig.~3(g) elucidates the potential versus time curves, which originate from the continuous current pulse of duration ($\tau$) 5 min followed by a relaxation time of 30 min for both positive and negative current, and the response is consistent with the galvanostatic charge-discharge plateaus in Fig.~2(a). The diffusion coefficient of Na$^{+}$ can be calculated using the Fick's second law of diffusion and the linear relationship between E and $\tau$$^{1/2}$ (see Fig.~S3(c) of ref.~\cite{SIfile}), as given below \cite{LiNJC21, RuiEA10}: 
\begin{equation}
D_{Na^{+}}= \frac{4}{\pi \tau}\left[\frac{m_B V_M}{M_B A}\right]^2\left[\frac{\Delta E_s}{\Delta E_t}\right]^2;\tau=L^{2}/D_{Na^{+}}
\end{equation}
where, {\it m}$_{B}$ is the mass loading of the material (g), {\it M}$_{B}$ is the molecular weight (g mol$^{-1}$), {\it V}$_{M}$ is the molar volume (cm$^{3}$ mol$^{-1}$), {\it A} is the active surface area between the electrode and electrolyte (here, for simplification we have considered the geometrical surface area, i.e., 1.130 cm$^{2}$), {\it L} is the thickness of the electrode, $\Delta$ {\it E}$_{t}$ represents the change in potential during the application of current pulse, and $\Delta$ {\it E}$_{s}$ attributes the potential difference between the equilibrium states or steady state voltage. A better representation of all the parameters can be seen in Fig.~3(h) by using a single titration curve. The values of the diffusion coefficient are plotted in a logarithmic scale with respect to the change in potential of the cell during both charging and discharging process, as shown in Fig.~3(i). The diffusivity of Na-ion varies in the range of 10$^{-8.5}$ to 10$^{-10.5}$ cm$^{2}$s$^{-1}$ for charging/discharging process. Intriguingly, we are able to improve the Na-ion diffusion kinetics by developing the CNT-modified composite material of NaFe$_{1.6}$V$_{0.4}$(PO$_{4}$)(SO$_{4}$)$_{2}$, as compared to the values reported in ref.~\cite{EssehliJPS20}. 

\subsection{\noindent ~Temperature dependent electrochemical study:} 

In order to understand the active ion kinetics and the cell dynamics during different physical and chemical processes, we perform the EIS measurements in 25--55$^\circ$C  temperature range and extract the kinetic parameters involved in the SEI formation and charge transfer mechanics through detailed analysis. Note that the Na-ion diffusion can be interpreted as a four step process through Barsoukov's model \cite{BarsoukovSSI3}, where the ion first migrate through the liquid electrolyte and solid electrolyte interface (SEI) layer followed by a charge transfer across the SEI/electrode interface, and the final process involves the solid diffusion inside the active material. Fig.~4(a) shows the EIS spectra as the Nyquist plot at different temperatures with the equivalent circuit in the inset. We observe that the diameter of the semi-circles in the fitted Nyquist plots decreases with temperature. This indicates the lower resistances of SEI layer (R$_{SEI}$) and charge transfer process (R$_{ct}$), as more amount of self discharge occurs followed by sodium plating and dendrite formation at high temperatures \cite{KondouEC17}. We use the following Arrhenius equation derive the activation energy involved during the Na-ion transport through SEI layer and SEI/electrode interface  \cite{LiNC19, BeraJPCC20}:
 \begin{eqnarray}
 \sigma T = A exp(-E_{a}/K_{B}T)
 \end{eqnarray} 
where, {\it A} is the proportionality factor and can also be termed as frequency factor, {\it E}$_{a}$ is the activation energy, $\sigma$ is the ionic conductivity, {\it K}$_{B}$ is the Boltzmann constant and {\it T} is the absolute temperature. Here, $\sigma$ represents the multiplication product of cell constant (c) and reciprocal of resistance obtained from the fitted circuits. As the cell constant of the interface can not be obtained, the y-axis of Fig.~4(b) is denoted as log (R$^{-1}${\it T}) instead of log (cR$^{-1}${\it T}) \cite{BuscheNCH16}. The slope of log (R$^{-1}${\it T}) and 1000/{\it T} provide the activation energy for the respective chemical process \cite{ChoudhuryNC17} and we found the values 0.29~eV and 0.27~eV in case of charge transfer across SEI/electrode interface and the SEI layer, respectively, as shown in Fig.~4(b). In both the cases, the Na-ion transport face low activation energy due to the surface modification of the active material with CNT, which provides large active surface area with interconnected frame-work to increase the electronic conductivity of NFVPS material. Further, the studies of bode plot and Na-ion kinetics at different temperatures are provided in Figs.~S3(a, b) of ref.~\cite{SIfile}. 

Fig.~4(c) displays the CV curves measured at different temperatures between 25 and 55\degree C with a scan rate of 0.5 mVs$^{-1}$. Intriguingly, the oxidation and reduction peaks are almost at same potential and similar current values for all the temperatures. This indicates that the CNT modified cathode material is highly reversible for the applications in a wide range of temperature up to 50\degree C. However, slight change is observed in the peak intensity and the voltage values, which signifies the low stability of the electrode/electrolyte combination and large IR drop with some irreversible behavior, respectively, when measured at 55\degree C \cite{RuiEA10}. Further, we use a fresh half-cell and tested at a current rate of 5~C with 11 cycles at each temperature between 25 and 55\degree C, as shown in Fig.~4(d), which demonstrate the improvement in the specific capacity and stability at higher temperature. This is due the fast reaction kinetics and higher Na$^{+}$ diffusion, as both sodiation and SEI forming kinetics in the host material increase at high temperatures \cite{LiangAMI17}. Further we observe small fluctuations in values of specific capacity at 55\degree C, which might be due to some electrolyte degradation \cite{ZhuEM20}. It is likely that the SEI layer forms due to electrolyte degradation when the cell is cycled at high temperature and high-potential region, which can result in some irreversible capacity and low Coulombic efficiency at 55\degree C. Also, there can be some overcharge issues in the initial cycles at each temperature that may result in some fluctuations in the Coulombic efficiency \cite{LiangJMCA19, ZhaoCEJ20}. Moreover, in Fig.~4(f) we present the long cycling measurement of NFVPS@CNT cathode at a high current rate of 5~C, which shows high stability with 80\% capacity retention up to 100 cycles at 45\degree C. The faster Na-ion kinetics and smaller charge transfer resistance at 45\degree C provides smaller IR drop, which improves the specific capacity and cycling stability. This may be possible due to enhanced conductivity of the NFVPS materials with CNT composite, which helps to prevent the undesirable side reactions between the electrode and electrolyte during long cycling at high temperatures \cite{Li_JMCA19}. 

\begin{figure*}
\includegraphics[width=6.8in]{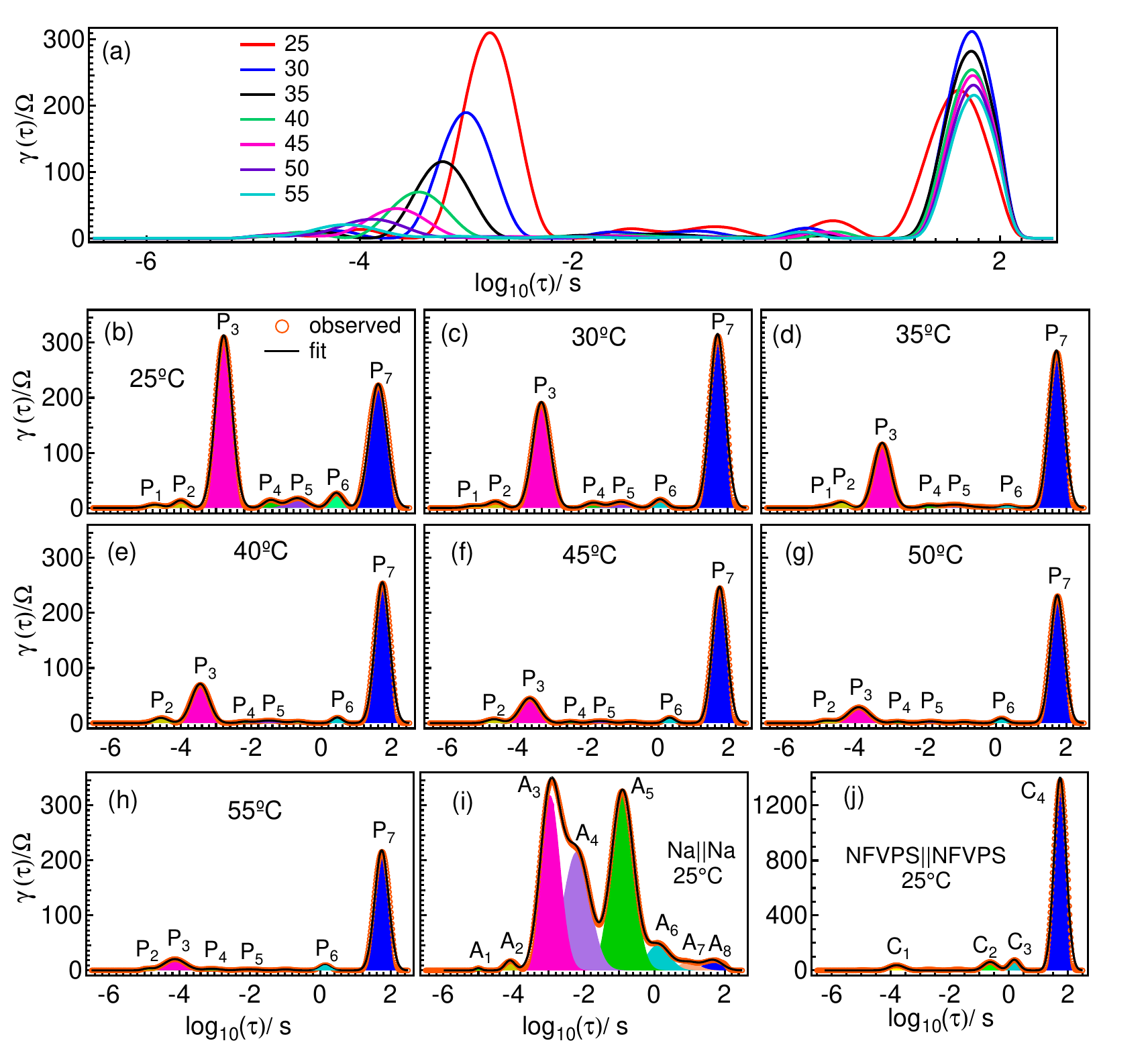}
\caption {{\bf The Distribution of Relaxation Time (DRT) analysis of NaFe$_{1.6}$V$_{0.4}$(PO$_{4}$)(SO$_{4}$)$_{2}$@CNT cathode:} the simulated EIS spectra (a) compared at different temperatures, (b) 25$^\circ$C, (c) 30$^\circ$C, (d) 35$^\circ$C, (e) 40$^\circ$C,(f) 45$^\circ$C, (g) 50$^\circ$C and (h) 55$^\circ$C, (i) the Na\textbar\textbar Na symmetric cell, and (j) the NFVPS\textbar\textbar NFVPS symmetric cell at 25$^\circ$C. The peaks marked P1 to P7 in (b--h) are corresponding to the contact and charge-transfer resistances as well as the diffusion process inside the cell. The peaks marked with A1 to A8 and C1 to C4 are corresponding to the anodic and cathodic contributions of different physical processes.}
 \label{EIS}
\end{figure*}

\subsection{\noindent ~Distribution of Relaxation Time analysis:} 

The EIS fitting model includes more number of circuit elements and overlap semicircles in the same frequency domain, which make its analysis more difficult for the electrochemical processes. Therefore, a more reliable mathematical tool named DRT (Distribution of Relaxation Time) is used to analyse the cell behavior in the time domain by measuring the large and small polarization effects that occurred in the similar frequency domain \cite{MeloEA21, ZhouJPS19-1, SoniESM22, DiGiuseppeEA20, ZhouJPS19}. In this method, an infinite Voigt circuit with series connection of parallel RQ elements is used to fit the impedance spectra \cite{SoniESM22} where the RQ determines a time constant, $\tau$ as following: 
\begin{equation}
\tau = (RQ)^{1/\alpha}    
\label{8} 
\end{equation}
where, {\it R} is a resistor, {\it Q} is the constant phase element (CPE) and $\alpha$ is a number between 0 to 1. The DRT tool provides an access to visualize individual processes of the system and the kinetic parameters involved in the particular process. Furthermore, this tool attributes higher resolution than the conventional EIS representation through Nyquist or Bode plots. We can also get quantitative information regarding the reaction kinetics by using the peak area, height and the position. For the DRT analysis, the measurement must be high quality and should satisfy time invariance. Therefore, Kramers-Kronig criterium is used to check the data validity. The link between DRT and EIS is established through the following equation developed by Fuoss and Kirkwood \cite{DiGiuseppeEA20}:
\begin{equation}
Z(\omega) = R_{s} + R_{p}\int_{-\infty}^{\infty}\frac{\gamma(\tau)}{1+j\omega\tau}dln\tau
\end{equation}
where, {\it R}$_{s}$ is the series resistance (also high frequency intercept), {\it R}$_{p}$ is the polarization resistance (low frequency intercept). In the DRT method, $\gamma(\tau)$ is a normalized function, which satisfy the following equation:
\begin{equation}
\int_{-\infty}^{\infty}\gamma(\tau)dln\tau = 1
\end{equation}
where, $\tau$ = 1/2$\pi$ f = 1/$\omega$. On subject to the above condition, the area under the curve of the following equation results the total polarization resistance of the system.
\begin{equation}
\int_{-\infty}^{\infty}\frac{\gamma(\tau)}{1+j\omega\tau}dln\tau
\end{equation}

For example, if we consider i$^{th}$ peak, then the time constant $\tau$$_{i}$ represents the position of maximum and the area of the peak defines the value of resistance R$_{i}$. So, we can calculate the value of capacitance as follows \cite{IurilliIEEE22}:
\begin{equation}
C_{i} = \tau_{i}/R_{i}
\end{equation}
The authenticity of the impedance curves is checked through Kramer-Kronig model using LIN-KK software, as shown in Fig.~4 of ref.~\cite{SIfile}. 
\begin{figure*}
\includegraphics[width=6.9in]{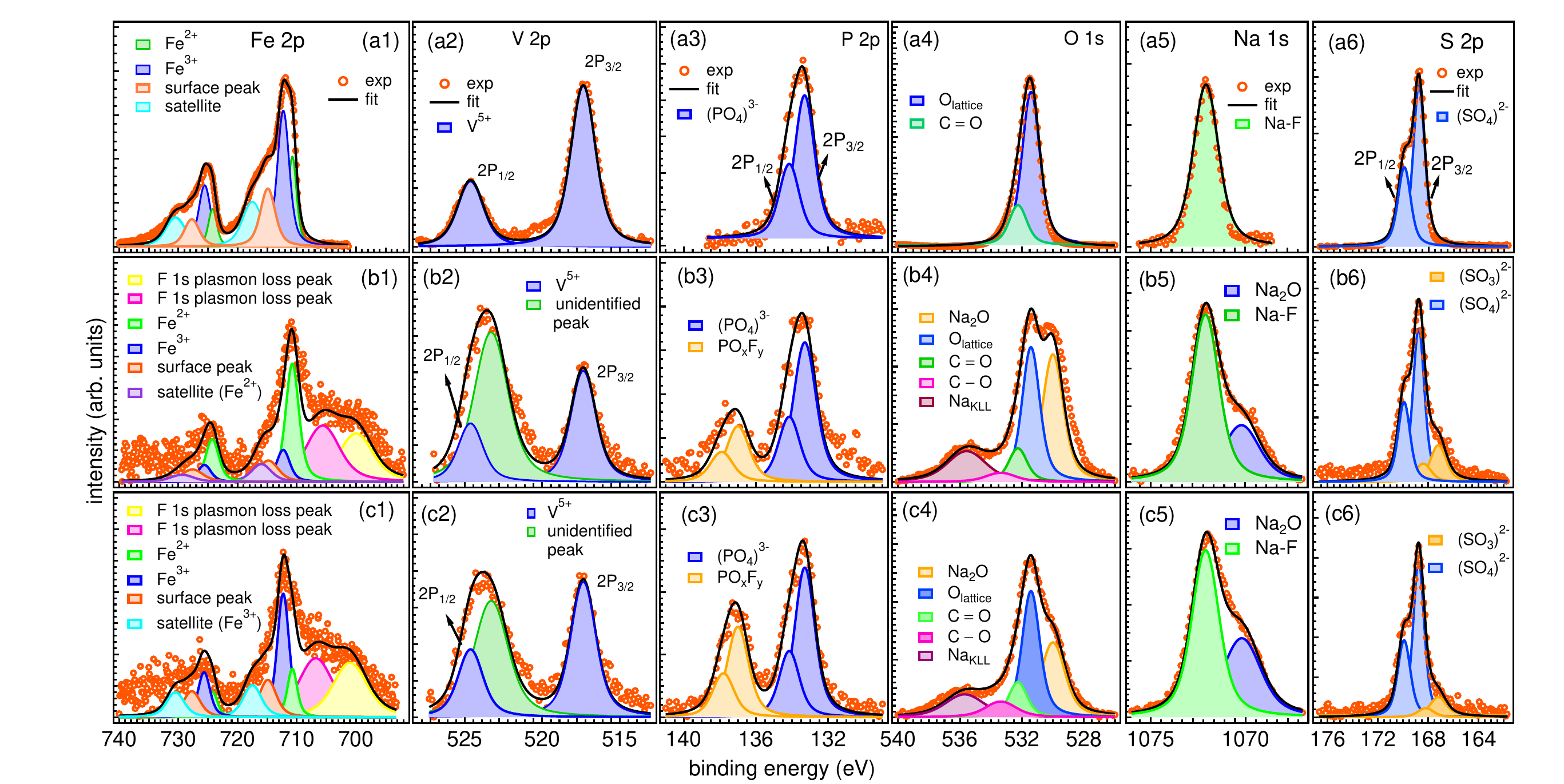}
\caption{{\bf The photoemission spectra of NaFe$_{1.6}$V$_{0.4}$(PO$_{4}$)(SO$_{4}$)$_{2}$@CNT cathode material:} The core-levels Fe 2$p$, V 2$p$, P 2$p$, O 1$s$, Na 1$s$ and S 2$p$ of (a1--a6) pristine sample, (b1--b6) in fully discharged and (c1--c6) fully charged states.}
 \label{XPS}
\end{figure*}
We note that the DRT profiles generated from the EIS data in Fig.~4(a) show different local maxima (peaks) in the distribution function $\gamma(\tau)$, see Fig.~5(a). The area of these peaks interpret the total polarization resistance ({\it R}$_{p}$) and the shift in their position depicts the contribution of different physical processes inside the cell. It has been reported that the time constant domain $\le$10$^{-3}$ s represents the resistance between particle-particle and particle-current collector, between 10$^{-3}$--10$^{-2}$ s and 10$^{-2}$--10$^{-1}$ s indicate the active ion transportation through the SEI layer and charge transfer resistance of the active ion at cathode and anode interface, respectively; and $\ge$10$^{-1}$ s attributes to the diffusion process of active ion in the bulk electrode \cite{ZhouJPS19,  ChenJPS21, ManikandanJPS17, SabetJPS20}.

The fitted DRT profiles at different temperatures in Figs.~5(b--h) show seven local maxima (marked by P1--P7), where each peak can be differentiated according to the frequency range of the EIS spectrum. In Fig.~5(b), the DRT peaks P1 and P2 positioned at 8.82 ms and 18.16 ms, which determine the contact resistance indicating the high-frequency region of the semi-circle. Further, the peaks P3, P4, P5, P6 at time constants 61.90, 235, 500 and 658 ms constitute the mid-frequency range of the semicircle representing the charge transfer resistance at SEI and cathode-electrolyte/anode-electrolyte interface. At last the peak P7 at 5~s in the low frequency region depicts the diffusion process inside the solid. In order to differentiate the cathodic and anodic origin of these peaks, we tested the symmetric Na\textbar\textbar Na and NFVPS@CNT\textbar\textbar NFVPS@CNT cells \cite{SoniESM22, WangET22}, as shown in Figs.~5(i, j), respectively. Fig.~5(i) shows a strong peak A3 in the region of 10$^{-4}$--10$^{-2}$~s, whereas in Fig.~5(j) the peak C1 in the same region is very weak in intensity. This indicates that the peak P3 in mid-frequency region shows dominant anodic contribution due to the the charge transfer resistance at anode/electrolyte interface (SEI). By comparing the intensity of peaks A3 and C4 around 5~s in Figs.~5(i, j), respectively, we can conclude that the peak P7 in low frequency region resulted from a dominant cathodic contribution. Interestingly, we observe that all the peaks P1--P7 in Figs.~5(b--h) are sensitive to the change in temperature, where a significant decrease in intensity of peak P3 is visible at high temperatures. This suggests that the large resistance at the anode/electrolyte interface (SEI) at around room temperature is due to the lower ionic conductivity of the electrolyte, which slows down the charge-transfer processes \cite{IurilliIEEE22}. On the other hand, the relative intensity of peak P7 first increase at 30$^\circ$C, implying higher activation barrier for diffusion of active ions, then it decreases gradually indicating higher diffusivity of ions at high temperatures. The DRT analysis is consistent with the observed temperature dependent electrochemical behavior in Fig.~4, and found to be very useful to understand the kinetics of electrode materials at different temperatures by separating between cathodic and anodic contributions. The DRT can also be used to study the surface reactivity, performance limiting mechanisms and cell aging \cite{ZhouJPS19, DiGiuseppeEA20, ZhouJPS19-1, MohsinJPS22}. 

\subsection{\noindent ~Ex-situ photoemission spectroscopy study:} 

\begin{figure*}
\includegraphics[width=6.7 in]{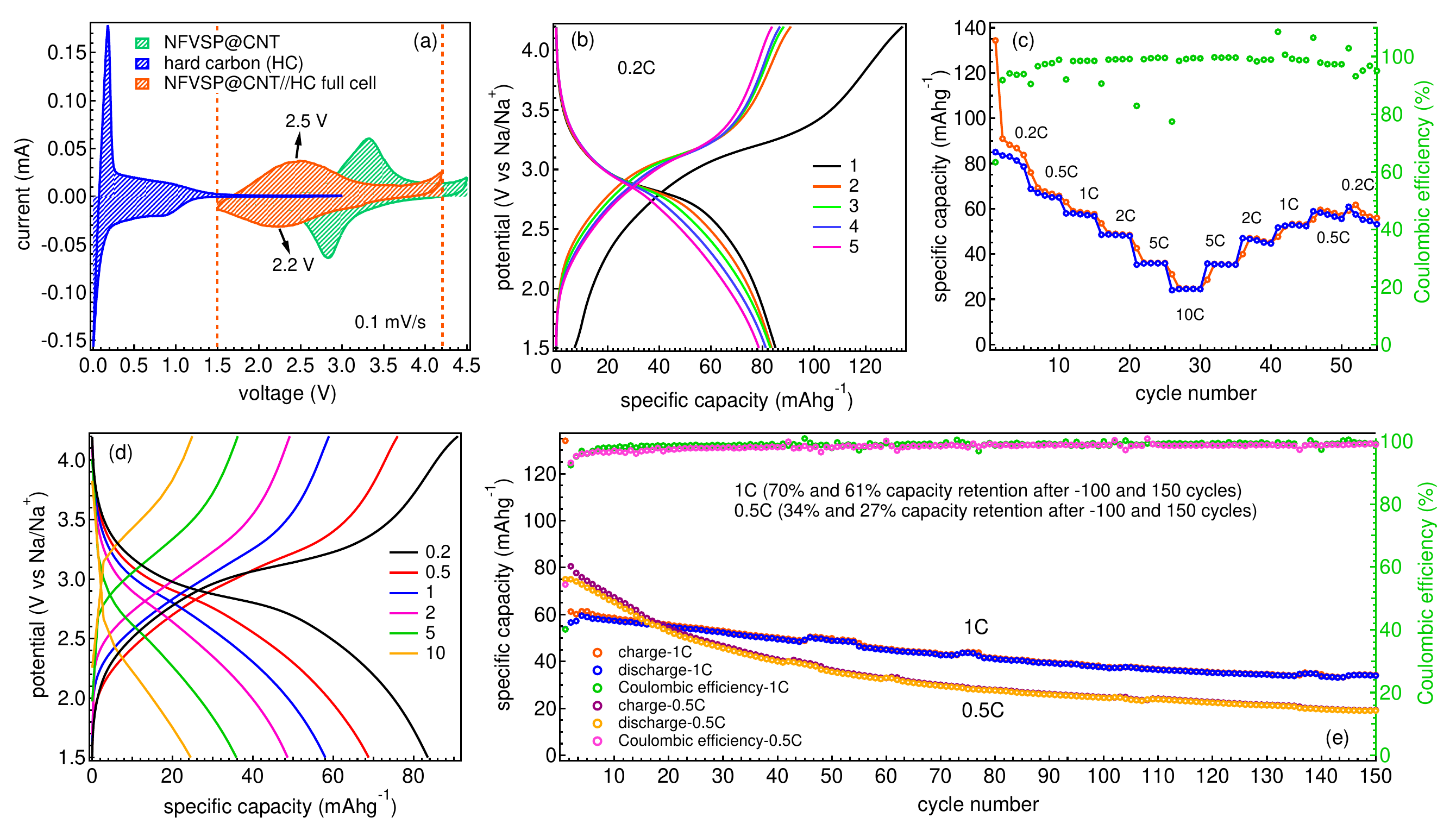}
\caption{{\bf Electrochemical performance of a full cell battery (NFVPS@CNT\textbar\textbar HC) in a voltage window of 1.5--4.2~V:} (a) The CV curves of HC as anode, NFVPS@CNT as cathode and NFVPS@CNT//HC full cell at a scan rate of 0.1 mVs$^{-1}$, (b) the GCD plots of initial 5 cycles at 0.2~C, (c) the rate capability measurements at different current rates for 5 cycles each with Coulombic efficiency, (d) the corresponding galvanostatic discharge-charge voltage profiles of 2$^{nd}$ cycle at different current rates from 0.2 to 10 C of (c). The specific capacity and Coulombic efficiency (\%) in (e) measured at current rates of 1~C and 0.5~C up to 150 cycles.}
\label{FC}
\end{figure*}

Finally, we use x-ray photoelectron spectroscopy (XPS) to investigate the elemental oxidation state of pristine NFVPS@CNT material as well as in the fully charge and discharge states after Na-ion insertion/de-insertion. In Fig.~6, we present the core-level spectra of Fe 2$p$, V 2$p$, P 2$p$, O 1$s$, Na 1$s$, and S 2$p$ for the pristine sample (a1--a6), after the first electrochemical discharge (b1--b6) and after first charge (c1--c6), respectively. The Fe 2$p$ core-level split into two spin-orbit components namely 2$p_{3/2}$ and 2$p_{1/2}$, where the deconvolution with Voigt function confirm the presence of both Fe$^{2+}$ and Fe$^{3+}$ oxidation states in the ratio of 32\% and 68\%, respectively. For the pristine sample, the binding energy (BE) values of the Fe$^{3+}$ peaks are found to be 712.3~eV (2$p_{3/2}$) and 725.6~eV (2$p_{1/2}$), respectively; whereas, for the Fe$^{2+}$, the respective peaks are situated at 710.7~eV and 724.2~eV \cite{CarreraMSEB21}. Also, the presence of surface peak at 715~eV corresponds to the Fe$^{3+}$ ions in the sublayers and the peak at 717.6~eV represents the shake-up satellite of Fe$^{3+}$ ions \cite{CarreraMSEB21}. In the discharge process, the Fe$^{3+}$ is expected to reduce to Fe$^{2+}$ and during the charging process it should oxidize back to Fe$^{3+}$, which found to be consistent with the observation of dominant Fe$^{2+}$ ($\approx$78\%) in the fully discharged state [Fig.~6(b1)] and a dominant Fe$^{3+}$ contribution of 76\% in fully charged state [Fig.~6(c1)]. We also observe the evolution of two new peaks at 701 and 706.7~eV, marked as the plasmon loss features of F 1$s$ \cite{GrosvenorSIA04} in Figs.~6(b1, c1), which are probably due to the degradation of NaPF$_{6}$ electrolyte and formation of NaF at electrode/electrolyte interface during charging \cite{BasaJMST20, BeletskiiENG19}, as also marked in Fig.~S6 of ref.~\cite{SIfile}.  

The V 2$p$ core-level spectra of the pristine, discharge and charge states are shown in Figs.~6(a2, b2, c2), respectively. In Fig. 6(a2), the spin-orbit components at 517.3 eV (2p$_{3/2}$) and 524.6 eV (2p$_{1/2}$) indicate the V$^{5+}$ oxidation state in the pristine sample \cite{WagnerNIST03}. In Figs.~6(b2, c2) we find no shift in the V 2$p$ peak positions confirming the V in 5+ state for charge/discharge states, which is consistent as the V is not expected to participate in the electrochemical process as also evident in Figs.~2(a, b). Also, for both the cases (charge/discharge) we observe a strong feature at 523.3~eV [Figs.~6(b2, c2)] whose origin seems to be controversial in the literature; therefore, we marked this as unidentified peak. It has been reported that this peak might be due to Na KLL Auger electrons, plasmon loss peaks, and O 1s photoelectrons from the interphase and SEI layers \cite{QuerelCM23}. The P 2$p$ core-level spectra in Figs.~6(a3, b3, c3) show spin-orbit peaks at 133.3 and 134.1~eV corresponding to the 2$p_{3/2}$ and 2$p_{1/2}$, respectively, consistent with chemical environment for P element in (PO$_{4}$)$^{3-}$ anions \cite{YanAEM19}. However, in Figs.~6(b3, c3) we observe extra peaks with spin-orbit splitting around 137~eV (2$p_{3/2}$) and 138~eV (2$p_{1/2}$) in discharge/charge states. These suggest a P-F linkage or the generation of PO$_{x}$F$_{y}$ products from NaPF$_{6}$ degradation during electrochemical process \cite{YanAEM19}. 

The O 1$s$ spectra in Figs.~6(a4, b4, c4) exhibit a consistent peak at 531.4 eV indicating oxygen atoms in the (PO$_{4}$)$^{3-}$ and (SO$_{4}$)$^{2-}$ groups or lattice oxygen \cite{YanAEM19}. The weak broad characteristics at 532.2 and 533.4 ~eV indicate oxygenated carbon at the surface \cite{YanAEM19}. We observe that during charge and discharge, the oxygenated species and Na$_{2}$O production at the electrode surface cause a new peak at 530~eV \cite{ChenJMCA20} as well as Na KLL auger peak at 535.7~eV \cite{DubeyAEM21, LiJAC24}, see Figs.~6(b4, c4). The BE value of the Na 1$s$ core-levels spectra is found to be at 1072.1~eV for all the three samples [see Figs.~6(a5, b5, c5)]. This peak can be assigned as a signal of NaF formation, a useful SEI component, which helps to improve the ionic conductivity and other mechanical properties \cite{KalapsazovaJMCA14}. In Figs.~6(b5, c5), the asymmetry towards lower BE side fits a peak at 1070.2~eV, suggesting the presence of Na$_{2}$O in the charge/discharge states \cite{ChenJMCA20}. The S 2$p$ core level peaks in Figs.~6(a6, b6, c6) are observed at 168.7 (2$p_{3/2}$) and 169.9 eV (2$p_{1/2}$) \cite{AndreuAMI15}. In case of the charge and discharge states, a more reduced form of S as (SO$_{3}$)$^{2-}$ peaks are visible at 167.2 (2$p_{3/2}$) and 168.4~eV (2$p_{1/2}$) \cite{AndreuAMI15}. In case of the charge and discharge states, a more reduced form of S as (SO$_{3}$)$^{2-}$ peaks are visible at 167.2 (2$p_{3/2}$) and 168.4~eV (2$p_{1/2}$) due to the interaction between the edge sulphur and oxygen from the surface \cite{AndreuAMI15}. Also, it is possible that weak features of thiosulfonate (SO$_{3}$)$^{2-}$ species may overlap for charging state owing to the oxidation of (SO$_{3}$)$^{2-}$ to (SO$_{4}$)$^{2-}$ \cite{AndreuAMI15}. 

\subsection{\noindent ~Full cell study:} 

Lastly we fabricate and test the full cell battery by coupling NFVPS@CNT as cathode and hard carbon (HC) as anode. The CV curves of NFVPS@CNT\textbar\textbar HC full cell in comparison with HC anode and NFVPS@CNT cathode at a scan rate of 0.1 mVs$^{-1}$ are shown in Fig.~7(a). There is one pair of redox peaks 2.5/2.25~V for the full cell, which are closest to the NFVPS@CNT-Na half cell with a lower redox polarization of 0.25~V. Fig.~7(b) shows the galvanostatic charge-discharge profiles of the full-cell in a voltage window of 1.5-4.2~V where we observe a maximum specific discharge capacity of 85 mAhg$^{-1}$ at 0.2~C and maintains 93\% of initial capacity retention after 5 cycles. Notably, considering the nominal voltage of 2.7~V, the specific energy density of the full cell is evaluated to be around 155 Whkg$^{-1}$ \cite{SIfile} considering the active material masses of both negative and positive electrodes, which found to be excellent for Na-ion battery system \cite{CaoAFM21}. Moreover, the rate capability data in Fig.~7(c) depicts excellent electrochemical performance of the full cell battery up to high current rate of 10~C. Also, the same cell was cycled back to 0.2~C which found to retain 78\% of the initial specific capacity. Fig.~7(d) shows the corresponding GCD profiles at different current rates in respective 2$^{nd}$ cycle. Further, the long cycling measurements in Fig.~7(e) at current rates of 1~C and 0.5~C exhibit 70\% and 34\% capacity retentions after 100 cycles, respectively. The poor stability at slower current rate is the occurrence of interfacial reactions due to relatively large exposure time of electrolytes at high potential around 4.2~V \cite{ChandelDT22, PatiJMCA22}. There is also inhibition of some reactions and structural evolution at higher current rates, which helps to provide high-capacity retention as compared to the lower current rates \cite{ChandelDT22, PatiJMCA22}. The better stability at higher current rates indicates the high-rate capability nature of the full cell battery. Furthermore, the full cells exhibit faster capacity decay than half cells owing to the limited sodium inventory in full cells, as it is partially consumed on the hard carbon surface through irreversible process during SEI formation \cite{SmithBS23}.  

\section{\noindent ~Summary and Conclusions:} 

In summary, we demonstrated the mixed polyanionic NFVPS@CNT composite as fast electronic and ionic conduction owing to the 3D interconnected channels, high specific capacity, thermal stability and reversibility as cathode material in sodium-ion batteries (SIBs). The cathode delivers a high discharge specific capacity of 104 mAhg$^{-1}$ at 0.1~C with an average working voltage of $\sim$3~V. Furthermore, the rate capability measurements upto a high current rate of 25~C, reveal the fast charging activity of the material and it retains 95\% of initial capacity after returning back to 0.1~C. Notably, the long cycling measurement at different current rate up to 10~C provides 57\% and 36\% capacity retention after 1000 and 2000 cycles, respectively. The evaluated diffusion coefficient of the electrode material through CV and GITT falls in the range of 10$^{-8}$ to 10$^{-11}$ cm$^{2}$s$^{-1}$, which is considered as a fast sodium-ion transport. We have employed DRT analysis to differentiate between the individual physical processes occurring inside the half-cell at different temperatures in the frequency domain. Intriguingly, the temperature dependent electrochemical investigation reveal the thermal stability and significant capacity retention of 80\% at 5~C rate at 45$^\circ$C up to 100 cycles. Moreover, {\it ex-situ} XRD, FE-SEM and XPS results of the cycled materials confirm the stable structural, morphological and electronic properties. The full cell battery shows $\sim$155 Wh kg$^{-1}$ (85~mAhg$^{-1}$ and 2.7~V) specific energy at 0.2~C (considering the active material weight of both the electrodes). Our study provide a new path to develop 3D interconnected phosphate--sulfate frameworks in CNT modified NFVPS electrode to improve the energy density of SIBs. 

\section{\noindent ~Conflicts of interest:} 

The authors declare no competing financial interest. 

\section{\noindent ~Acknowledgments:} 

The research facilities for this work on sodium-ion battery project are financially supported by Department of Science and Technology, Government of India through ``DST--IIT Delhi Energy Storage Platform on Batteries" (project no. DST/TMD/MECSP/2K17/07) and SERB--DST through core research grant (file no.: CRG/2020/003436). JP thanks UGC for the fellowship. We thank IIT Delhi for providing characterization facilities (XRD and Raman scattering at the physics department, and XPS, FE-SEM and HR-TEM at CRF). \\


\begin{thebibliography}{99}

\bibitem{YabuuchiCR14} N. Yabuuchi, K. Kubota, M. Dahbi, S. Komaba, Research development on sodium-ion batteries, Chem. Rev. 114 (2014) 11636--11682.

\bibitem{ArmandNature08} M. Armand, J. -M. Tarascon, Building better batteries, Nature 451 (2008) 652--657.

\bibitem{VaalmaNRM18} C. Vaalma, D. Buchholz, M. Weil, S. Passerini, A cost and resource analysis of sodium-ion batteries, Nat. Rev. Mater. 3 (2018) 18013.

\bibitem{LiNE22} Y. Li, Q. Zhou, S. Weng, F. Ding, X. Qi, J. Lu, Y. Li, X. Zhang, X. Rong, Y. Lu, X. Wang, R. Xiao, H. Li, X. Huang, L. Chen, Y.-S. Hu,  Interfacial engineering to achieve an energy density of over 200 Wh Kg$^{-1}$ in sodium batteries, Nat. Energy 7 (2022) 511.

\bibitem{GaoESM20} R.-M. Gao, Z.-J. Zheng, P.-F. Wang, C.-Y. Wang, H. Ye, F.-F. Cao, Recent Advances and Prospects of Layered Transition Metal Oxide Cathodes for Sodium-Ion Batteries, Energy Storage Mater. 30 (2020) 9.

\bibitem{YouAEM18} Y. You, A. Manthiram, Progress in high‐voltage cathode materials for rechargeable sodium‐ion batteries, Adv. Energy Mater. 8 (2018) 1701785.

\bibitem{LiAFM20} H. Li, M. Xu, Z. Zhang, Y. Lai, J. Ma, Engineering of Polyanion Type Cathode Materials for Sodium‐Ion Batteries: Toward Higher Energy/Power Density, Adv Funct Materials 30 (2020) 2000473. 

\bibitem{BarpandaNC14} P. Barpanda, G. Oyama, S. Nishimura, S. -C. Chung, A. Yamada, A 3.8-V earth-abundant sodium battery electrode, Nat. Commun. 5 (2014) 4358.

\bibitem{SapraWEE21} S. K. Sapra, J. Pati, P. K. Dwivedi, S. Basu, J. Chang, R. S. Dhaka, A comprehensive review on recent advances of polyanionic cathode materials in Na‐ion batteries for cost effective energy storage applications, WIREs Energy \& Environment 10 (2021) e400.

\bibitem{LiueS24} C. Liu, K. Chen, H. Xiong, A. Zhao, H. Zhang, Q. Li, X. Ai, H. Yang, Y. Fang, Y. Cao, A Novel Na$_{8}$Fe$_{5}$(SO$_{4}$)$_{9}$@rGO Cathode Material with High Rate Capability and Ultra-Long Lifespan for Low-Cost Sodium-Ion Batteries, eScience 4 (2024) 100186. 

\bibitem{TianJMCA22} K. Tian, H. He, X. Li, D. Wang, Z. Wang, R. Zheng, H. Sun, Y. Liu, Q. Wang, Boosting Electrochemical Reaction and Suppressing Phase Transition with a High-Entropy O3-Type Layered Oxide for Sodium-Ion Batteries, J. Mater. Chem. A 10 (2022) 14943.

\bibitem{DangASS23} Y. Dang, Z. Xu, H. Yang, K. Tian, Z. Wang, R. Zheng, H. Sun, Y. Liu, and D. Wang, Designing Water/Air-Stable Co-Free High-Entropy Oxide Cathodes with Suppressed Irreversible Phase Transition for Sodium-Ion Batteries, Appl. Surf. Sci. 636  (2023) 157856.

\bibitem{LinCEC22} X. Lin, X. Ren, L. Cong, Y. Liu, X. Xiang, Reversible Multi‐Electron Reaction Mechanism of Sodium Vanadium/Manganese Phosphate Cathode for Enhanced Na‐Storage Capability, ChemElectroChem 9 (2022) e202200669.

\bibitem{KimJACS12} H. Kim, I. Park, D. -H. Seo, S. Lee, S. -W. Kim, W. J. Kwon, Y. -U. Park, C. S. Kim, S. Jeon, K. Kang, New iron-Based mixed-polyanion cathodes for lithium and sodium rechargeable batteries: combined first principles calculations and experimental study, J. Am. Chem. Soc. 134 (2012) 10369--10372.

\bibitem{ChenCM13} H. Chen, Q. Hao, O. Zivkovic, G. Hautier, L. -S. Du, Y. Tang, Y. -Y. Hu, X. Ma, C. P. Grey, G. Ceder, Sidorenkite (Na$_{3}$MnPO$_{4}$CO$_{3}$): a new intercalation cathode material for Na-Ion batteries, Chem. Mater. 25 (2013) 2777--2786.

\bibitem{NoseJPS13} M. Nose, H. Nakayama, K. Nobuhara, H. Yamaguchi, S. Nakanishi, H. Iba, Na$_{4}$Co$_{3}$(PO$_{4}$)$_{2}$P$_{2}$O$_{7}$: A novel storage material for sodium-ion batteries, J. Power Sources 234 (2013) 175--179.

\bibitem{DwivediACSAEM21} P. K. Dwivedi, S. K. Sapra, J. Pati, R. S. Dhaka, Na$_{4}$Co$_{3}$(PO$_{4}$)$_{2}$P$_{2}$O$_{7}$/NC composite as a negative electrode for sodium-ion batteries, ACS Appl. Energy Mater. 4 (2021) 8076--8084. 

\bibitem{ShivaEES16} K. Shiva, P. Singh, W. Zhou, J. B. Goodenough, NaFe$_{2}$PO$_{4}$(SO$_{4}$)$_{2}$: a potential cathode for a Na-ion battery, Energy Environ. Sci. 9 (2016) 3103--3106.

\bibitem{YahiaJPS18} H. Ben Yahia, R. Essehli, R. Amin, K. Boulahya, T. Okumura, I. Belharouak, Sodium intercalation in the phosphosulfate cathode NaFe$_{2}$PO$_{4}$(SO$_{4}$)$_{2}$, J. Power Sources 382 (2018) 144--151.

\bibitem{LiNJC21} S. -F. Li, X. -K. Hou, Z. -Y. Gu, Y. -F. Meng, C. -D. Zhao, H. -X. Zhanga, X. -L. Wu, Sponge-like NaFe$_{2}$PO$_{4}$(SO$_{4}$)$_{2}$@rGO as a high-performance cathode material for sodium-ion batteries, New J. Chem. 45 (2021) 4854--4859.

\bibitem{EssehliJPS20} R. Essehli, A. Alkhateeb, A. Mahmoud, F. Boschini, H. Ben Yahia, R. Amin, I. Belharouak,  Optimization of the compositions of polyanionic sodium-ion battery cathode NaFe$_{2-x}$V$_{x}$PO$_{4}$(SO$_{4}$)$_{2}$, J. Power Sources 469 (2020) 228417.

\bibitem{ChenNC19} M. Chen, W. Hua, J. Xiao, D. Cortie, W. Chen, E. Wang, Z. Hu, Q. Gu, X. Wang, S. Indris, S. -L. Chou, S. -X. Dou,  NASICON-type air-stable and all-climate cathode for sodium-ion batteries with low cost and high-power density, Nat Commun. 10 (2019) 1480 .

\bibitem{LanBS21} T. Lan, Q. Ma, C. Tsai, F. Tietz, O. Guillon, Ionic Conductivity of Na$_{3}$V$_{2}$P$_{3}$O$_{12}$ as a Function of Electrochemical Potential and its Impact on Battery Performance, Batteries \& Supercaps 4 (2021) 479--484.

\bibitem{LiCAEJ23} X. Li, Y. Meng, D. Xiao, Three‐dimensional holey graphene modified Na$_{4}$Fe$_{3}$(PO$_{4}$)$_{2}$P$_{2}$O$_{7}$/C as a high‐performance cathode for rechargeable sodium‐ion batteries, Chem. Eur. J. 29 (2023) e202203381.

\bibitem{LiangAMI17} X. Liang, X. Ou, F. Zheng, Q. Pan, X. Xiong, R. Hu, C. Yang, M. Liu, Surface modification of Na$_{3}$V$_{2}$(PO$_{4}$)$_{3}$ by nitrogen and sulfur dual-doped carbon layer with advanced sodium storage property, ACS Appl. Mater. Interfaces 9 (2017) 13151--13162. 

\bibitem{IurilliIEEE22} P. Iurilli, C. Brivio, R. Carrillo, V. Wood,  EIS2MOD: a DRT-based modeling framework for Li-ion cells, IEEE Trans. on Ind. Applicat. 58 (2022) 1429--1439.

\bibitem{SIfile} see Supplementary information for additional analysis of HR-TEM, FESEM-EDX, post-cycling treatments, some temperature dependent electrochemical investigations and full cell studies. 

\bibitem{QiuIEE19} J. Qiu, X. Li, X. Qi, Raman spectroscopic investigation of sulfates using mosaic grating spatial heterodyne Raman spectrometer, IEEE Photon. J. 11 (2019) 1--12.

\bibitem{AngelAS12} S. M. Angel, N. R. Gomer, S. K. Sharma, C. McKay, Remote raman spectroscopy for planetary exploration: a review, Appl. Spectrosc. 66 (2012) 137--150.

\bibitem{MarkevichJPS11} E. Markevich, R. Sharabi, O. Haik, V. Borgel, G. Salitra, D. Aurbach, G. Semrau, M. A. Schmidt, N. Schall, C. Stinner, Raman spectroscopy of carbon-coated LiCoPO$_{4}$ and LiFePO$_{4}$ olivines, J. Power Sources 196 (2011) 6433--6439.

\bibitem{WuNano13} J. Wu, G. K. P. Dathar, C. Sun, M. G. Theivanayagam, D. Applestone, A. G. Dylla, A. Manthiram, G. Henkelman, J. B. Goodenough, K. J. Stevenson, In situ Raman spectroscopy of LiFePO$_{4}$ : size and morphology dependence during charge and self-discharge, Nanotechnology 24 (2013) 424009.

\bibitem{DifiJPCC15} S. Difi, I. Saadoune, M. T. Sougrati, R. Hakkou, K. Edstrom, P.-E. Lippens, Mechanisms and performances of Na$_{1.5}$Fe$_{0.5}$Ti$_{1.5}$(PO$_{4}$)$_{3}$ /C composite as electrode material for Na-ion batteries, J. Phys. Chem. C 119 (2015) 25220--25234.

\bibitem{CriadoFP19} A. Criado, P. Lavela, G. F. Ortiz, J. L. Tirado, S. Gzouli, Z. Edfouf, C. Pérez-Vicente, CTAB-assisted synthesis of C@Na$_{3}$V$_{2}$(PO$_{4}$)$_{2}$F$_{3}$ with optimized morphology for application as cathode material for Na-ion batteries, Front. Phys. 7 (2019) 207.

\bibitem{GaoJMC17} C. Gao, J. Zhou, G. Liu, L. Wang, Microwave-assisted synthesis and surface decoration of LiFePO4 hexagonal nanoplates for lithium-ion batteries with excellent electrochemical performance, J. Mater. Sci. 52 (2017) 1590. 

\bibitem{ChoiJN13} Y. C. Choi, K. -I. Min, M. S. Jeong, Novel method of evaluating the purity of multiwall carbon nanotubes using Raman spectroscopy, J. Nanomater. 2013 (2013) 1--6.

\bibitem{MelvinJMC14} G. J. H. Melvin, Q. -Q. Ni, Y. Suzuki, T. Natsuki, Microwave-absorbing properties of silver nanoparticle/carbon nanotube hybrid nanocomposites, J. Mater. Sci. 49 (2014) 5199--5207.

\bibitem{ChandelDT22} S. Chandel, Zulkifli, J. Kim, A. K. Rai, Effect of vanadium doping on the electrochemical performances of sodium titanate anode for sodium ion battery application, Dalton Trans. 51 (2022) 11797--11805.

\bibitem{GuanMAT21} Z. Guan, Z. Zhang, B. Du, Z. Peng, A Non-Flammable Zwitterionic Ionic Liquid/Ethylene Carbonate Mixed Electrolyte for Lithium-Ion Battery with Enhanced Safety, Materials 14 (2021). 4225.

\bibitem{ChoiPCCP18} D. Choi, J. Kang, J. Park, B. Han, First-Principles Study on Thermodynamic Stability of the Hybrid Interfacial Structure of LiMn2O4 Cathode and Carbonate Electrolyte in Li-Ion Batteries, Phys. Chem. Chem. Phys. 20 (2018) 11592.

\bibitem{PatiJMCA22} J. Pati, H. Raj, S. K. Sapra, A. Dhaka, A. K. Bera, S. M. Yusuf and R. S. Dhaka, Unraveling the diffusion kinetics of honeycomb structured Na$_{2}$Ni$_{2}$TeO$_{6}$ as a high-potential and stable electrode for sodium-ion batteries, J. Mater. Chem. A 10 (2022) 15460--15473.

\bibitem{SinghCEJ23} M. K. Singh, J. Pati, D. Seth, J. Prasad, M. Agarwal, M. A. Haider, J.-K. Chang, R. S. Dhaka, Diffusion mechanism and electrochemical investigation of 1T phase Al–MoS$_{2}$@rGO nano-composite as a high-performance anode for sodium-ion batteries, Chem. Eng. J. 454 (2023) 140140.

\bibitem{PuACI21} X. Pu, D. Zhao, C. Fu, Z. Chen, S. Cao, C. Wang, Y. Cao, Understanding and Calibration of Charge Storage Mechanism in Cyclic Voltammetry Curves, Angew Chem Int Ed 60 (2021) 21310 .

\bibitem{RuiEA10} X. H. Rui, N. Ding, J. Liu, C. Li, C. H. Chen, Electrochim. Acta 55 (2010) 2384--2390.

\bibitem{BarsoukovSSI3} E. Barsoukov, Comparison of kinetic properties of LiCoO$_{2}$ and LiTi$_{0.05}$Mg$_{0.05}$Ni$_{0.7}$Co$_{0.2}$O$_{2}$ by impedance spectroscopy, Solid State Ion. 161 (2003) 19--29.

\bibitem{KondouEC17} H. Kondou, J. Kim, H. Watanabe, Thermal analysis on Na plating in sodium ion battery, Electrochemistry 85 (2017) 647--649.

\bibitem{LiNC19} K. Li, J. Zhang, D. Lin, D.-W. Wang, B. Li, W. Lv, S. Sun, Y.-B. He, F. Kang, Q.-H. Yang, L. Zhou, T.-Y. Zhang, Evolution of the electrochemical interface in sodium ion batteries with ether electrolytes, Nat. Commun. 10 (2019) 725.

\bibitem{BeraJPCC20} A. K. Bera, S. M. Yusuf, Temperature-dependent Na-ion conduction and its pathways in the crystal structure of the layered battery material Na$_{2}$Ni$_{2}$TeO$_{6}$, J. Phys. Chem. C 124 (2020) 4421--4429.

\bibitem{BuscheNCH16} M. R. Busche, T. Drossel, T. Leichtweiss, D. A. Weber, M. Falk, M. Schneider, M.-L. Reich, H. Sommer, P. Adelhelm, J. Janek, Dynamic formation of a solid-liquid electrolyte interphase and its consequences for hybrid-battery concepts, Nat. Chem. 8 (2016) 426--434.

\bibitem{ChoudhuryNC17} S. Choudhury, S. Wei, Y. Ozhabes, D. Gunceler, M. J. Zachman, Z. Tu, J. H. Shin, P. Nath, A. Agrawal, L. F. Kourkoutis, T. A. Arias, L. A. Archer, Designing solid-liquid interphases for sodium batteries, Nat Commun. 8 (2017) 898.

\bibitem{ZhuEM20} X. Zhu, L. Wang, Advances in materials for all‐climate sodium‐ion batteries, Eco Mat 2 (2020) e12043.

\bibitem{LiangJMCA19} L. Liang, W. Zhang, D. K. Denis, J. Zhang, L. Hou, Y. Liu, C. Yuan, Comparative investigations of high-rate NaCrO 2 cathodes towards wide-temperature-tolerant pouch-type Na-ion batteries from -15 to 55 $\degree$C: nanowires vs. bulk, J. Mater. Chem. A 7 (2019) 11915--11927.

\bibitem{ZhaoCEJ20} X. Zhao, Z. Gu, W. Li, X. Yang, J. Guo, X. Wu, Temperature‐dependent electrochemical properties and electrode kinetics of Na$_{3}$V$_{2}$(PO$_{4}$)$_{2}$O$_{2}$F cathode for sodium‐ion batteries with high energy density, Chem. Eur. J. 26 (2020) 7823--7830.

\bibitem{Li_JMCA19} S. Li, X. Song, X. Kuai, W. Zhu, K. Tian, X. Li, M. Chen, S. Chou, J. Zhao, L. Gao, A Nano-architectured Na$_6$Fe$_5$(SO$_4$)$_8$/CNTs Cathode for Building a Low-Cost 3.6 V Sodium-Ion Full Battery with Superior Sodium Storage, J. Mater. Chem. A 7 (2019) 14656.

\bibitem{MeloEA21} B. M. G. Melo, F. J. A. Loureiro, D. P. Fagg, L. C. Costa, M. P. F. Graca, DFRT to EIS: an easy approach to verify the consistency of a DFRT generated from an impedance spectrum, Electrochim. Acta 366 (2021) 137429.

\bibitem{ZhouJPS19-1} X. Zhou, Z. Pan, X. Han, L. Lu, M. Ouyang, An easy-to-implement multi-point impedance technique for monitoring aging of lithium ion batteries, J. Power Sources 417 (2019) 188--192.

\bibitem{SoniESM22} R. Soni, J. B. Robinson, P. R. Shearing, D. J. L. Brett, A. J. E. Rettie, T. S. Miller, Lithium-sulfur battery diagnostics through distribution of relaxation times analysis, Energy Stor. Mater. 51 (2022) 97--107.

\bibitem{DiGiuseppeEA20} G. DiGiuseppe, A. Hunter, F. Zhu, Combined equivalent circuits and distribution of relaxation times analysis and interfacial effects of (La$_{0.60}$Sr$_{0.40}$)$_{0.95}$Co$_{0.20}$Fe$_{0.80}$O$_{3-x}$ cathodes, Electrochim. Acta 350 (2020) 136252.

\bibitem{ZhouJPS19} X. Zhou, J. Huang, Z. Pan, M. Ouyang, Impedance characterization of lithium-ion batteries aging under high-temperature cycling: Importance of electrolyte-phase diffusion, J. Power Sources 426 (2019) 216--222.

\bibitem{ChenJPS21} X. Chen, L. Li, M. Liu, T. Huang, A. Yu, Detection of lithium plating in lithium-ion batteries by distribution of relaxation times, J. Power Sources 496 (2021) 229867.

\bibitem{ManikandanJPS17} B. Manikandan, V. Ramar, C. Yap, P. Balaya, Investigation of physico-chemical processes in lithium-ion batteries by deconvolution of electrochemical impedance spectra, J. Power Sources 361 (2017) 300--309.

\bibitem{SabetJPS20} P. Shafiei Sabet, A. J. Warnecke, F. Meier, H. Witzenhausen, E. Martinez-Laserna, D. U. Sauer, Non-invasive yet separate investigation of anode/cathode degradation of lithium-ion batteries (nickel–cobalt–manganese vs. graphite) due to accelerated aging, J. Power Sources 449 (2020) 227369.

\bibitem{WangET22} X. Wang, C. Chen, S. Wu, H. Zheng, Y. Chen, H. Liu, Y. Wu, H. Duan, High‐rate and long‐life Au nanorods/LiFePO$_{4}$ composite cathode for lithium‐ion batteries, Energy Tech. 10 (2022) 2100841.

\bibitem{MohsinJPS22} U. Mohsin, L. Schneider, M. Häringer, C. Ziebert, M. Rohde, W. Bauer, H. Ehrenberg, H. J. Seifert, Heat Generation and Degradation Mechanisms Studied on Na$_3$V$_2$(PO$_4$)$_3$/C Positive Electrode Material in Full Pouch / Coin Cell Assembly, J. Power Sources 545 (2022) 231901. 

\bibitem{CarreraMSEB21} K. Carrera, V. Huerta, V. Orozco, J. Matutes, P. Fernández, O.A. Graeve, M. Herrera, Formation of vacancy point-defects in hydroxyapatite nanobelts by selective incorporation of Fe3+ ions in Ca(II) sites. A CL and XPS study, Mater. Sci. Eng. B 271 (2021) 115308.

\bibitem{GrosvenorSIA04} A. P. Grosvenor, B. A. Kobe, M. C. Biesinger, N. S. McIntyre, Investigation of multiplet splitting of Fe 2p XPS spectra and bonding in iron compounds, Surf. Interface Anal. 36 (2004) 1564--1574.

\bibitem{BasaJMST20} A. Basa, S. Wojtulewski, B. Kalska-Szostko, M. Perkowski, E. Gonzalo, O. Chernyayeva, A. Kuhn, F. G-. Alvarado, Carbon coating of air-sensitive insulating transition metal fluorides: An example study on $\alpha$-Li$_{3}$FeF$_{6}$ high-performance cathode for lithium ion batteries, J. Mater. Sci. Technol. 55 (2020) 107--115.

\bibitem{BeletskiiENG19} E. V. Beletskii, E. V. Alekseeva, D. V. Spiridonova, A. N. Yankin, O. V. Levin, Overcharge cycling effect on the surface layers and crystalline structure of LiFePO$_{4}$ cathodes of Li-ion batteries, Energies 12 (2019) 4652.

\bibitem{WagnerNIST03} C. D. Wagner, A. V. Naumkin, A. Kraut-Vass, J. W. Allison, C. J. Powell, J. R . Jr. Rumble, NIST Standard Reference Database 20 (2003) Version 3.4 (web version) (http:/srdata.nist.gov/xps/).

\bibitem{QuerelCM23} E. Quérel, N. J. Williams, I. D. Seymour, S. J. Skinner, A. Aguadero, Operando Characterization and Theoretical Modeling of Metal|Electrolyte Interphase Growth Kinetics in Solid-State Batteries. Part I: Experiments, Chem. Mater. 35 (2023) 853--862.

\bibitem{YanAEM19} G. Yan, K. Reeves, D. Foix, Z. Li, C. Cometto, S. Mariyappan, M. Salanne, J. Tarascon, A newelectrolyte formulation for securing high temperature cycling and storage performances of Na‐ion batteries, Adv. Energy Mater. 9 (2019) 1901431.

\bibitem{ChenJMCA20} Q. Chen, H. He, Z. Hou, W. Zhuang, T. Zhang, Z. Sun, L. Huang, Building an artificial solid electrolyte interphase with high-uniformity and fast ion diffusion for ultralong-life sodium metal anodes, J. Mater. Chem. A 8 (2020) 16232--16237. 

\bibitem{DubeyAEM21} B. P. Dubey, A. Vinodhkumar, A. Sahoo, V. Thangadurai, Y. Sharma, Microstructural tuning of solid electrolyte Na$_{3}$Zr$_{2}$Si$_{2}$PO$_{12}$ by polymer-assisted solution synthesis method and its effect on ionic conductivity and dielectric properties, ACS Appl. Energy Mater. 4 (2021) 5475--5485. 

\bibitem{LiJAC24} T. Li, M. Lu, Y. Zhang, X. Xiang, S. Liu, and C. Chen, Structural Evolution and Redox Chemistry of Robust Ternary Layered Oxide Cathode for Sodium-Ion Batteries, J. Alloys Compd. 978 (2024) 173459.

\bibitem{LiJAC24} T. Li, M. Lu, Y. Zhang, X. Xiang, S. Liu, and C. Chen, Structural Evolution and Redox Chemistry of Robust Ternary Layered Oxide Cathode for Sodium-Ion Batteries, J. Alloys Compd. 978 (2024) 173459.

\bibitem{KalapsazovaJMCA14} M. Kalapsazova, R. Stoyanova, E. Zhecheva, G. Tyuliev, D. Nihtianova, Sodium deficient nickel–manganese oxides as intercalation electrodes in lithium ion batteries, J. Mater. Chem. A 2 (2014) 19383--19395.

\bibitem{AndreuAMI15} N. Andreu, D. Flahaut, R. Dedryvère, M. Minvielle, H. Martinez, D. Gonbeau, XPS investigation of surface reactivity of electrode materials: effect of the transition metal, ACS Appl. Mater. Interfaces 7 (2015) 6629--6636.

\bibitem{CaoAFM21} Y. Cao, X. Cao, X. Dong, X. Zhang, J. Xu, N. Wang, Y. Yang, C. Wang, Y. Liu, Y. Xia, All-Climate Iron-Based Sodium-Ion Full Cell for Energy Storage, Adv Funct Materials 31 (2021) 2102856.

\bibitem{SmithBS23} A. Smith, P. Stüble, L. Leuthner, A. Hofmann, F. Jeschull, and L. Mereacre, Potential, Limitations of Research Battery Cell Types for Electrochemical Data Acquisition, Batter. Supercaps 6 (2023) e202300080.

\end{thebibliography}
\end{document}